\documentclass[aps,nofootinbib,superscriptaddress,twocolumn]{revtex4-1}
\usepackage{graphicx}
\usepackage{amsmath,amssymb}
\usepackage{hyperref}
\usepackage{braket}
\usepackage{booktabs}
\usepackage{subfigure}
\usepackage{subfloat}
\usepackage{float}
\usepackage[usenames]{color}

\hypersetup{
    colorlinks=true,
    linkcolor=red,
    citecolor=blue,
}

\def\be{\begin{equation}}
\def\ee{\end{equation}}
\def\ba{\begin{eqnarray}}
\def\ea{\end{eqnarray}}

\begin{document}


\title{Constraints on Primordial Magnetic Fields from Planck combined with the South Pole Telescope CMB B-mode polarization measurements}

\author{Alex Zucca}
\affiliation{Department of Physics, Simon Fraser University, Burnaby, BC, V5A 1S6, Canada}

\author{Yun Li}
\affiliation{Department of Physics, Simon Fraser University, Burnaby, BC, V5A 1S6, Canada}

\author{Levon Pogosian}
\affiliation{Department of Physics, Simon Fraser University, Burnaby, BC, V5A 1S6, Canada}
\affiliation{Institute of Cosmology and Gravitation, University of Portsmouth, Portsmouth, PO1 3FX, UK}

\begin{abstract}
A primordial magnetic field (PMF) present before recombination can leave specific signatures on the cosmic microwave background (CMB) fluctuations. Of particular importance is its contribution to the B-mode polarization power spectrum. Indeed, vortical modes sourced by the PMF can dominate the B-mode power spectrum on small scales, as they survive damping up to a small fraction of the Silk length. Therefore, measurements of the B-mode polarization at high-$\ell$ , such as the one recently performed by the South Pole Telescope (\textsc{SPT}), have the potential to provide stringent constraints on the PMF. We use the publicly released SPT B-mode polarization spectrum, along with the temperature and polarization data from the \textsc{Planck} satellite, to derive constraints on the magnitude, the spectral index and the energy scale at which the PMF was generated. We find that, while \textsc{Planck} data constrains the magnetic amplitude to $B_{1 \, \text{Mpc}} < 3.3$ nG at 95\% confidence level (CL), the SPT measurement improves the constraint  to $B_{1 \, \text{Mpc}} < 1.5$ nG. The magnetic spectral index, $n_B$, and the time of the generation of the PMF are unconstrained. For a nearly scale-invariant PMF, predicted by simplest inflationary magnetogenesis models, the bound from \textsc{Planck+SPT} is $B_{1 \, \text{Mpc}} < 1.2$ nG at 95\% CL. For PMF with $n_B=2$, expected for fields generated in post-inflationary phase transitions, the 95\% CL bound is $B_{1 \, \text{Mpc}} < 0.002$ nG, corresponding to the magnetic fraction of the radiation density $\Omega_{B\gamma} < 10^{-3}$ or the effective field $B_{\rm eff} < 100$ nG. The patches for the Boltzmann code \textsc{CAMB} and the Markov Chain Monte Carlo engine \textsc{CosmoMC}, incorporating the PMF effects on CMB, are made publicly available.
\end{abstract}

\maketitle

\section{Introduction}

Magnetic fields exist in practically all gravitationally bound cosmic structures. They are seen in galaxies, with strengths of a few micro-Gauss ($\mu$G) and coherent over the extent of the galaxy, and in galaxy clusters, where they are of similar strength and extending well beyond the core regions \cite{Widrow:2002ud}. There is also preliminary evidence of magnetic fields coherent on mega-parsec (Mpc) scales permeating the inter-cluster space \cite{Neronov,Tashiro:2013ita}. The origin of the observed magnetic fields is not fully understood. The alignment of the galactic magnetic fields with the galactic disc planes suggests that they could be amplified via a dynamo process. However, the efficiency of the dynamo and the required strength of the initial seed field are still debated. Observations of $\mu$G strength fields in galaxies at redshifts $z>2$ \cite{Athreya:1998} add to the problem, as in such cases the dynamo would have only a short time to operate and would require a seed field as large as $10^{-11}$ Gauss \cite{Widrow:2011hs}. 

Mechanisms for generation of cosmic magnetic fields can be broadly divided into astrophysical and primordial. The proposed astrophysical scenarios include induction of fields at recombination \cite{Berezhiani:2003ik} or reionization \cite{Gnedin:2000ax}, either via the Biermann battery effect \cite{Biermann:1950} or photoionization \cite{Langer:2005gw,Ando:2010ry,Durrive:2015cja}, or the combination both \cite{Doi:2011qf}. In primordial scenarios, on the other hand, magnetic fields are produced in the very early universe, e.g. during inflation \cite{Turner:1987bw,Ratra:1991bn} or in phase transitions \cite{Vachaspati:1991nm}, and subsequently survive in a frozen-in state until the epoch of structure formation and collapse with the matter to seed the galactic fields. In particular, if the primordial magnetic field (PMF) was of nano-Gauss (nG) strength (in comoving units) and coherent over a comoving region of 1Mpc, there would be no need for a galactic dynamo, as the compression of the PMF within the proto-galactic halos would naturally produce $\mu$G strength fields. 

PMFs could be produced at several epochs in the early universe, including during and at the end of Inflation, as well as in the electroweak and the QCD phase transitions. While the resultant PMFs tend to be of very small strengths, the understanding of the details of the PMF generation and its subsequent evolution is by no means complete \cite{Durrer:2013pga}. Regardless of whether a PMF ends up being necessary for solving the galactic field problem, constraining it provides an observational handle for exploring fundamental physics in the very early universe. 

The cleanest window into the pre-recombination Universe is provided by the observations of the cosmic microwave background (CMB). A PMF contributes to CMB anisotropies through metric perturbation sourced by its stress-energy tensor and through the Lorentz force felt by the baryons in the plasma \cite{Mack:2001gc,Lewis:2004ef,Finelli:2008xh,Paoletti:2008ck,Shaw:2010}. In particular, the rotational (divergence-free) component of the velocity associated with the Lorentz force (the \emph{Alfv\'{e}n} mode) can cause dominant B-type polarization anisotropies that survive the small scale damping well below the Silk scale \cite{Subramanian:1997gi, Subramanian:1998fn, Seshadri:2000ky}. The PMF also modifies the ionization history of the universe \cite{Jedamzik:2013gua,Kunze:2014eka}, affecting the optical depth to the last scattering, although there are remaining uncertainties in modelling it \cite{Chluba:2015lpa}. Other effects of the PMF on CMB include Faraday rotation \cite{Kosowsky:1996yc,Kosowsky:2004zh,Pogosian:2011qv} and spectral distortions \cite{Kunze:2013uja} which are not well constrained by existing observations but are promising as future probes of the PMF.

The specific signatures imprinted by the PMF in the CMB spectra would carry valuable clues about its origin. For example, a PMF produced in a phase transition would have most of its power concentrated near the cutoff scale set by the plasma conductivity \cite{Jedamzik:1996wp,Durrer:2003ja,Jedamzik:2010cy} and could only affect the smallest CMB scales. On the other hand, the originally proposed inflationary models of magnetogenesis \cite{Turner:1987bw,Ratra:1991bn} predict a scale-invariant magnetic field, contributing to all observable CMB scales.

The analysis of the 2015 \textsc{Planck} data \cite{Adam:2015rua} performed in \cite{Ade:2015cva} limits the magnetic field strength smoothed over $1$Mpc to $B_{1{\rm Mpc}} < 4.4$ nG at the 95\% confidence level (CL), with the bound becoming stronger if a particular PMF spectrum is assumed. Prior to that, a comparable bound of $B_{1{\rm Mpc}} < 3.5$ nG at the 95\% was obtained in \cite{Paoletti:2012bb} from the combination of the 7-year \textsc{WMAP} data \cite{Larson:2010gs} and the high-$\ell$ temperature anisotropy spectrum from the South Pole Telescope (\textsc{SPT}) \cite{Keisler:2011aw}. More recently, a comparable bound of $B_{1{\rm Mpc}} < 3.9$ nG was derived by the \textsc{Polarbear} collaboration \cite{Ade:2015cao} based on their measurement of the B-mode polarization spectrum. This demonstrated the potential of the high $\ell$ B-mode measurements for constraining the PMF. Indeed, as we will show in this paper, adding the latest B-mode measurements by the \textsc{SPT} \cite{Keisler:2015hfa} significantly improves on the PMF bounds derived from the \textsc{Planck} data alone.

The paper is organized as follows. In Sec.~\ref{sect:PMFImpact} we discuss how non-helical primordial magnetic fields affect the CMB power spectra, in temperature and polarization. In Sec.~\ref{sect:DataAnalysis}, we derived constraints on the PMF from \textsc{Planck} and \textsc{SPT}. We conclude with a discussion in Sec.~\ref{sect:Discussion}.

\section{The PMF contributions to the CMB  spectra}
\label{sect:PMFImpact}

The impact of PMFs on CMB anisotropies has been studied in detail in \cite{Mack:2001gc,Lewis:2004ef, Finelli:2008xh, Paoletti:2008ck, Shaw:2010}. For the sake of completeness, we review the main points below. We will neglect effects associated with the Faraday rotation of CMB polarization as the \textsc{Planck} and the \textsc{SPT} data that we consider in this paper are unable to constrain them well.

We consider CMB fluctuations sourced by a stochastic magnetic field generated at a time $\tau_B$ in the early universe before the time  $\tau_{\nu}$ of neutrino decoupling. We restrict our treatment to linear order in perturbation theory at which the back-reaction of gravity on the PMF is ignored. We also assume that the unperturbed universe is spatially flat.

The ideal magnetohydrodynamic (MHD) limit holds to a good approximation on relevant scales in the highly conducting primeval plasma of the early universe. In this limit, the PMF is frozen in the plasma, and evolves according to $\mathbf{B}(\mathbf{x}, \tau) = \mathbf{B}(\mathbf{x}, \tau_0)/a^2(\tau)$ (see \cite{Subramanian:2015lua} for a review), where $\mathbf{B}$ is the magnetic field strength, $\tau_0$ denotes the present conformal time and $a$ is the scale factor normalized to $a(\tau_0) =1$. Conventionally, bounds on cosmological magnetic fields are quoted in terms of the ``comoving'' field strength $\mathbf{B}(\mathbf{x}, \tau_0)$. The electric field vanishes in the plasma in the MHD limit and, therefore, the energy-momentum tensor associated with the PMF can be written as \cite{Shaw:2010}
\begin{equation}
\label{eqn:PMFEnergyMomentumTensor}
\begin{split}
T^{\, \, 0}_{B \,0}(\mathbf{x}, \tau) & = - \frac{1}{8 \pi a^4} B^2(\mathbf{x}) \equiv -\rho_{\gamma} \Delta_B, \\
T^{\, \,i}_{B \,j}(\mathbf{x}, \tau) & = \frac{1}{4 \pi a^4} \biggl( \frac{1}{2} B^2(\mathbf{x}) \delta^i_j - B^{i}(\mathbf{x})B_j(\mathbf{x}) \biggr) \\
 & \equiv p_{\gamma}(\Delta_B \delta^i_j + \Pi^{\, \, i}_{B \, j}),
\end{split}
\end{equation}
where $B_i(\mathbf{x}) = B_i (\mathbf{x}, \tau_0)$, $\rho_\gamma$ and $p_\gamma=\rho_\gamma/3$ are the photon density and pressure, $\Delta_B$ is the magnetic contribution to the radiation density contrast and $\Pi^{\, \, i}_{B \, j}$ is the dimensionless anisotropic stress. Note that we assume absence of a homogeneous magnetic field at the background level, as it would break the isotropy of the Universe and has already been strongly constrained by CMB \cite{Barrow:1997mj}. The traceless symmetric tensor $\Pi_{B \,j}^{\,i}$ in Eq.~\eqref{eqn:PMFEnergyMomentumTensor}, can be decomposed in its scalar (S), vector (V) and tensor (T) components \cite{Kodama:1985bj}. We therefore expect PMF to source all metric perturbation modes, including gravitational waves. In Fourier space, the decomposition of $\Pi_{B \, j}^{\, i}$ reads
\begin{equation}
\label{eqn:PiBDecomp}
\Pi_{Bj}^{\,i} (\mathbf{k},\tau) e^{i\, \mathbf{k} \cdot \mathbf{x}}= \Pi_{Bj}^{\, i(S)} + \Pi_{Bj}^{\, i(V)} + \Pi_{Bj}^{\, i(T)},
\end{equation}
with the components given by
\begin{gather}
\label{eqn:ScalarDecomp}
\Pi_{B \,ij}^{(S)} = \Pi_B^{(0)}Q_{ij}^{(0)}, \\
\label{eqn:VectorDecomp}
\Pi_{B \,ij}^{(V)} = \Pi_B^{(+1)}Q_{ij}^{(+1)} +  \Pi_B^{(-1)}Q_{ij}^{(-1)} , \\
\label{eqn:TensorDecomp}
\Pi_{B \,ij}^{(T)} = \Pi_B^{(+2)}Q_{ij}^{(+2)} +  \Pi_B^{(-2)}Q_{ij}^{(-2)}, 
\end{gather}
where $Q_{ij}^{(0)} = -(\hat{k}_i \hat{k}_j - 1/3 \delta_{ij}) \exp{(i\, \mathbf{k} \cdot \mathbf{x})}$, $Q_{ij}^{(\pm 1)} = i \hat{k}_{( \, i} e^{\pm}_{j \, )} \exp{(i \, \mathbf{k} \cdot \mathbf{x})}$ and $Q_{ij}^{(\pm 2)} = e_{ij}^{(\pm 2)} \exp{(i \, \mathbf{k}\cdot\mathbf{x})}$ are, respectively, the scalar, vector and tensor harmonic functions \cite{Kodama:1985bj,Hu:1997hp}. Here, $\mathbf{e}^{(\pm)} = -i/\sqrt{2} (\mathbf{e}^1 \pm i \mathbf{e}^2)$, $\mathbf{e}^{1,2}$ are the unit vectors orthogonal to the wave vector $\mathbf{k}$ and $e_{ij}^{(\pm 2)} = \sqrt{3/2} e_i^{(\pm)} e_j^{(\pm)}$. 

We model the PMF as a statistically isotropic Gaussian distributed random field with the Fourier space two-point correlation given by
\begin{equation}
\label{eqn:MagneticPowerSpectrum}
\braket{B_i (\mathbf{k}) B_j^*(\mathbf{k}^{\prime})} = (2 \pi)^3 \delta^{(3)}(\mathbf{k} - \mathbf{k}^{\prime}) P_{ij} P_B(k), 
\end{equation}
where $P_{ij} = \delta_{ij} - \hat{k}_i \hat{k}_j$, and we neglect the helical contribution. In the special case of a maximally helical PMF, the effect of helicity on the parity-even CMB spectra (TT, TE, EE and BB) has been shown to be degenerate with a reduction in the overall normalization \cite{Kahniashvili:2005xe,Kunze:2011bp,Ballardini:2014jta}, leading to bounds on $B_{1{\rm Mpc}}$ that are weaker by about $25$\% \cite{Ade:2015cva}. Because of this, adding new unknown parameters to our analysis is not justified. We note, however, that helical fields also generate parity-odd spectra of EB and TB type \cite{Kahniashvili:2005xe,Kunze:2011bp,Ballardini:2014jta}, which could help in breaking the degeneracy with future CMB data.

\begin{figure*}[!htbp]
\centering
\subfigure{
\includegraphics[width=0.45\textwidth]{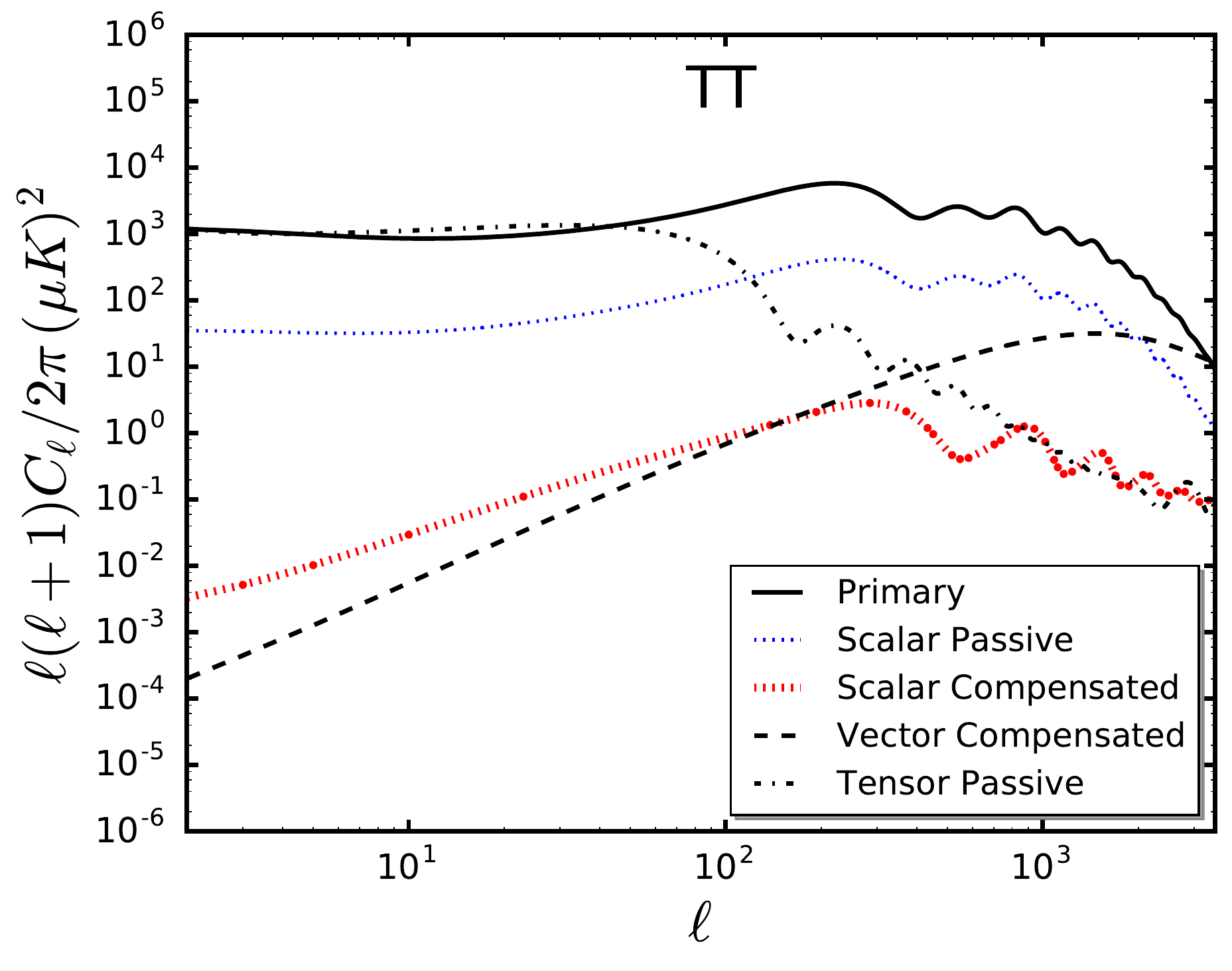}} \,
\subfigure{
\includegraphics[width=0.45\textwidth]{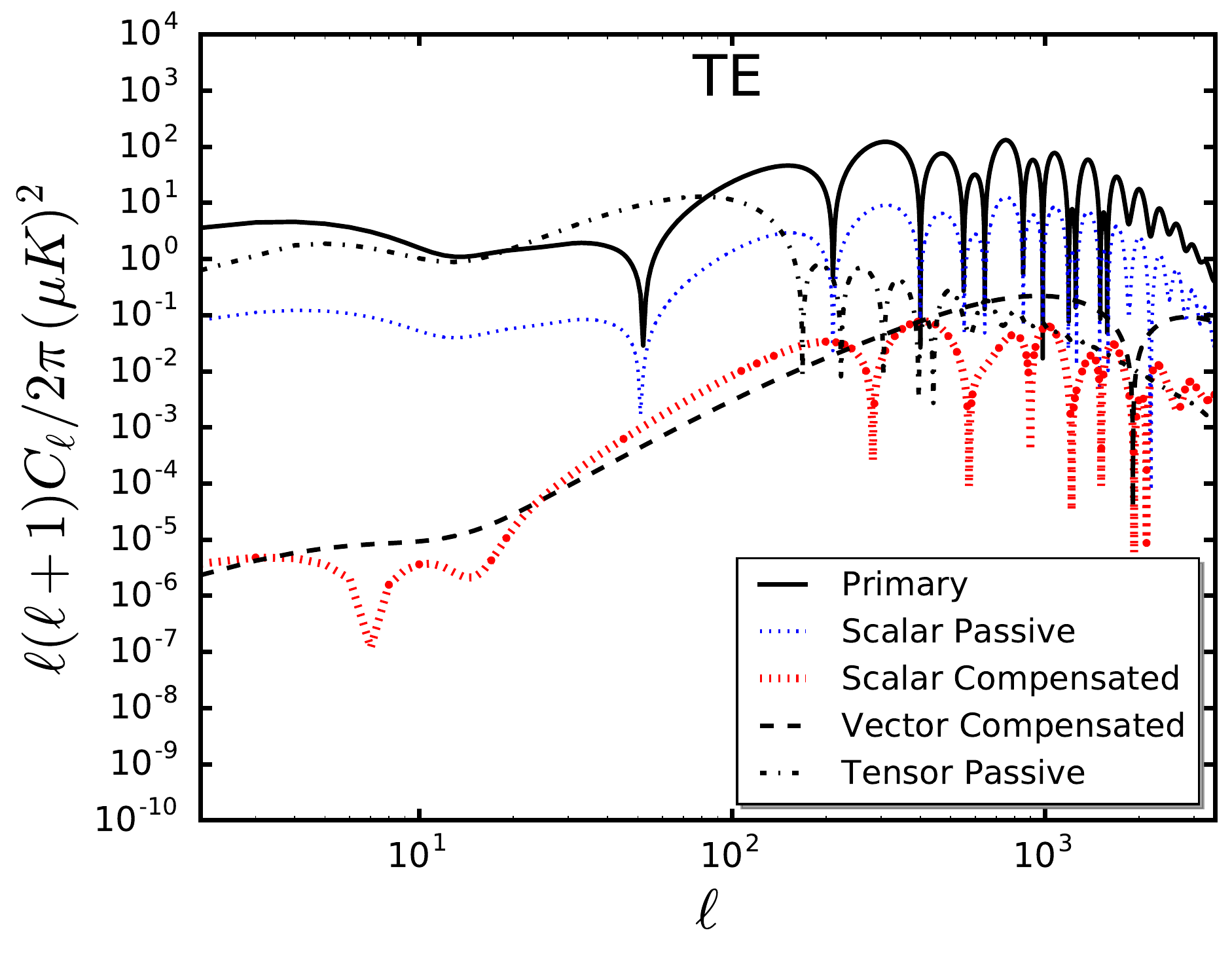}} \\
\subfigure{
\includegraphics[width=0.45\textwidth]{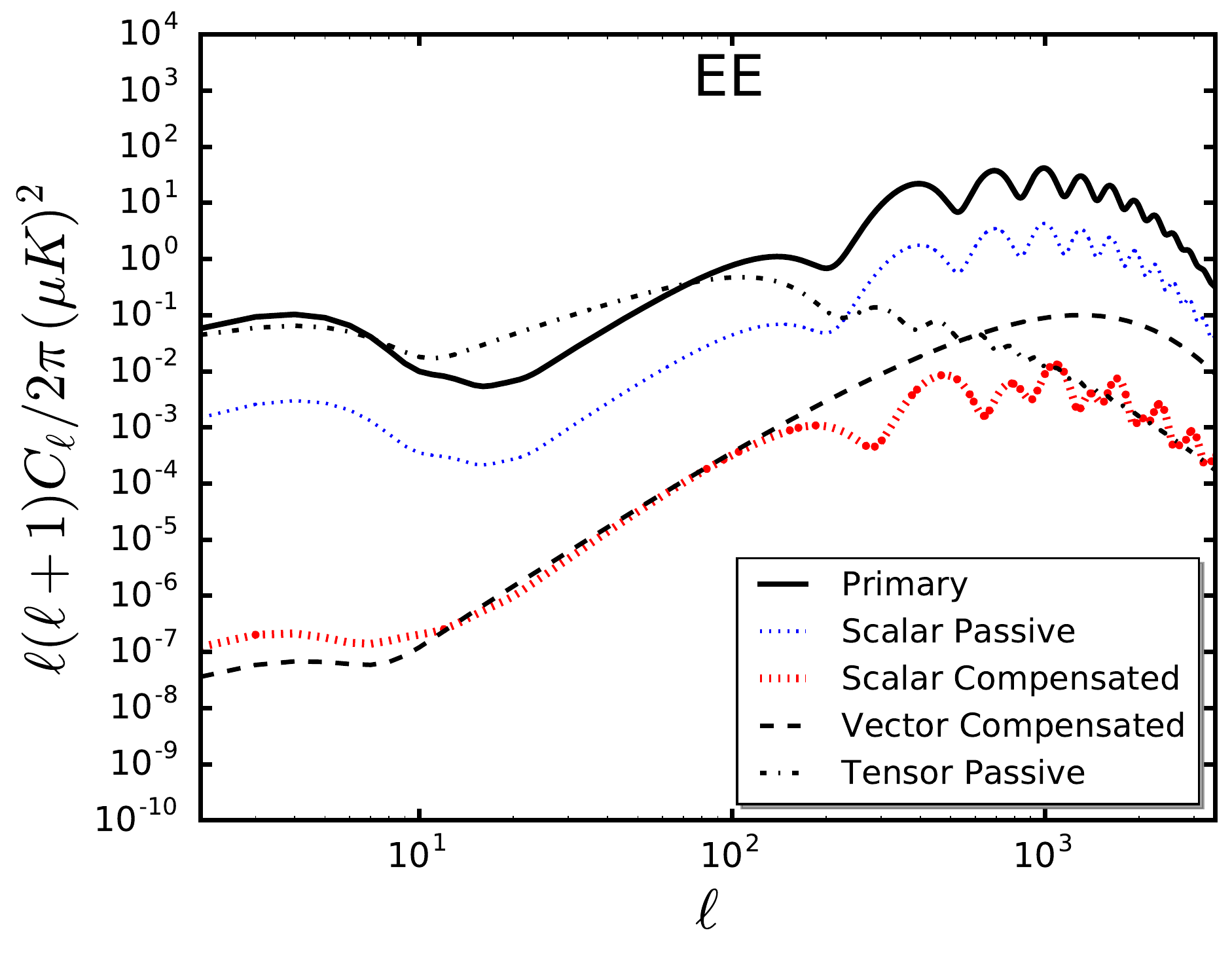}} \, 
\subfigure{
\includegraphics[width=0.45\textwidth]{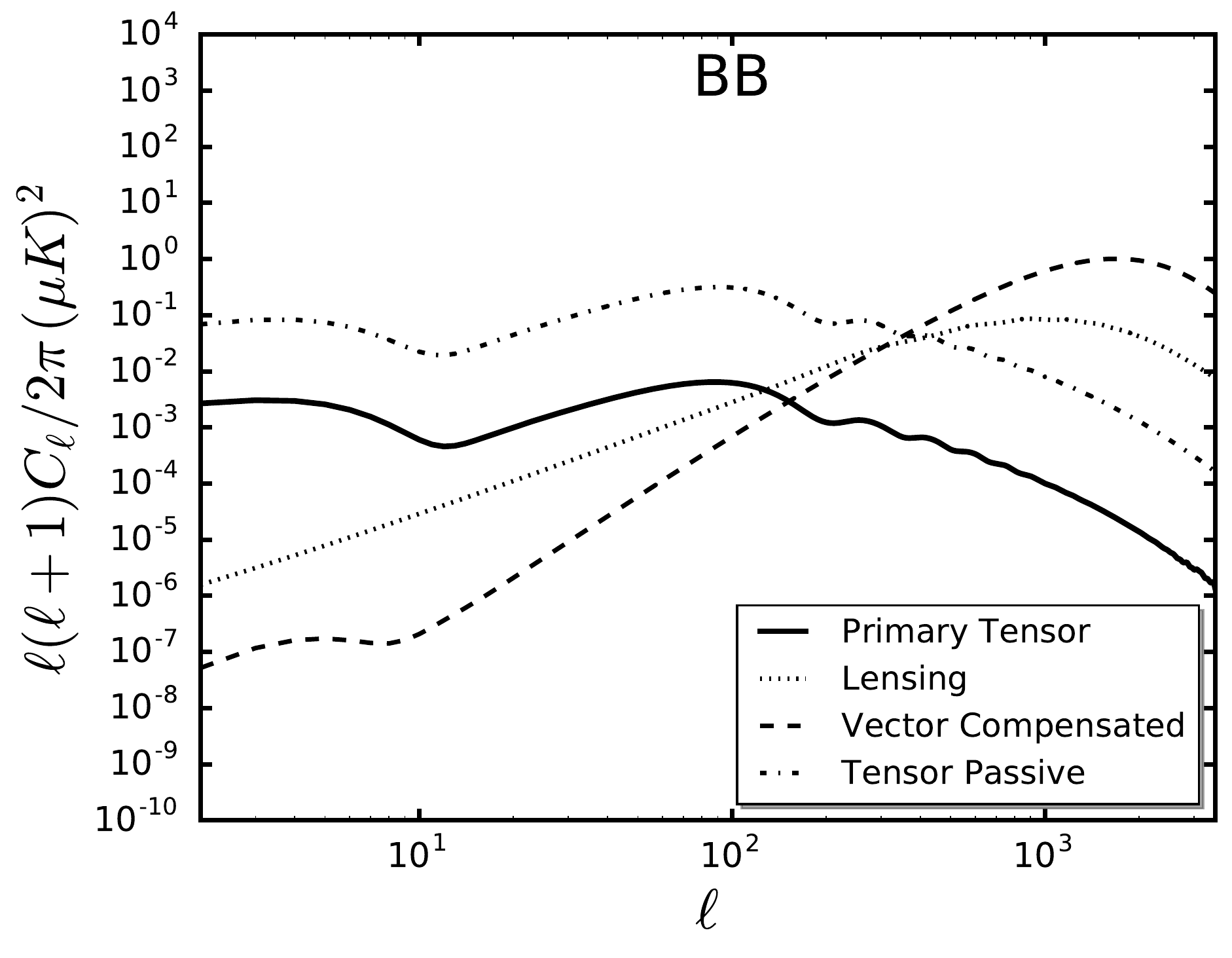}} 
\caption{\label{fig:MagneticModes} Contributions of relevant ``magnetic'' modes to the CMB temperature and polarization power spectra for a PMF with $B_{1\text{Mpc}}=4.5 \, \text{nG}$ and $n_B=-2.9$. For the passive modes, the time of the generation of the PMF is set at $\beta=\log_{10} (\tau_{\nu}/\tau_B) =17$. The cosmological parameters are set to $\omega_b = 0.0226$, $\omega_c = 0.112$, $T_{\rm CMB}=2.7255 \, \mathrm{K}$, $h=0.7$, $A_s = 2.1 \times 10^{-9}$, $n_s = 0.96$, $r=0.1$, $n_T = 1$.  The same parameters were used in Fig.~3 of \cite{Ade:2015cva}. Tensor compensated modes are negligible and are not shown.}
\end{figure*}

We take the magnetic power spectrum to be a power law,
\begin{equation}
P_B(k) = S_0 \, k^{n_B}
\end{equation}
for $k<k_D$, and zero otherwise. Here, $n_B$ is the spectral index, which depends on the mechanism that generates the PMF, and $2\pi/k_D$ is a damping scale below which magnetic fields dissipate due to radiation viscosity \cite{Jedamzik:1996wp, Subramanian:1997gi}. The damping scale depends on the amplitude and the spectral index of the PMF spectrum as \cite{Subramanian:1997gi,Mack:2001gc,Ade:2015cva}
\begin{equation}
{k_D\over \text{Mpc}^{-1}} = \biggl[ 5.5 \times 10^4 h \biggl( \frac{B_{\lambda}}{nG}\biggr)^{-2} \biggl( \frac{2 \pi}{\lambda / \text{Mpc}}  \biggr)^{n_B+3} \frac{\Omega_b h^2}{0.022} \biggr]^{\frac{1}{n_B+5}},
\end{equation}
where $\Omega_b$ is the baryon density fraction and $h$ is the reduced Hubble constant, $H_0 = 100 \, h \, \text{km}/(\text{s} \,\text{Mpc})$. We follow the convention in the literature and present the constraints on the PMF amplitude in terms of the smoothed amplitude $B_{\lambda}$, obtained by smoothing the magnetic energy density with a Gaussian filter over a comoving wavelength $\lambda$,
\begin{equation}
\begin{split}
B^2_{\lambda} = & \frac{1}{(2 \pi)^3} \int_{0}^{\infty} d^3 k \, e^{-k^2 \lambda^2} P_B(k)\\
= & \frac{2 S_0}{(2 \pi)^2} \frac{1}{\lambda^{n_B+3}}\Gamma \biggl( \frac{n_B+3}{2}\biggr),
\end{split}
\end{equation}
with $\lambda = 1 \, \text{Mpc}$. An alternative way of quantifying the amplitude of the PMF is in terms of its contribution to the radiation energy density, given by
\begin{equation}
\mathcal{E}_B = \frac{1}{(2 \pi)^3} \int_0^{k_D} dk \, k^2 P_B(k).
\end{equation}
One can then define the effective magnetic field strength as $B_{\text{eff}} = \sqrt{8 \pi \mathcal{E}_B}$, and the magnetic fraction of the radiation density as $\Omega_{B\gamma} = \mathcal{E}_B / \rho_\gamma$ where $\rho_\gamma$ is the total radiation energy density.  The relations between $B_{\text{eff}}$, $B_{\lambda}$ and $\Omega_{B\gamma}$ are given by \cite{Mack:2001gc,Pogosian:2011qv}
\begin{equation}
B_{\text{eff}} = \frac{B_{\lambda} (k_D \lambda)^{\frac{n_B+3}{2}}}{\sqrt{\Gamma ((n_B+5)/2)}} = 3.3 \times 10^{3} \sqrt{\Omega_{B\gamma}} \ {\rm nG}.
\label{eq:beff}
\end{equation}

Note that, since the stress energy of the PMF is quadratic in $\mathbf{B}$, the magnetic contributions to CMB anisotropies have a non-Gaussian statistics. The two point correlation functions of the magnetic energy momentum tensor components  $\Delta_B$ and $\Pi_B^{(0,\pm1,\pm2)}$ are derived from Eq.~\eqref{eqn:MagneticPowerSpectrum}. We show them in Appendix~\ref{sect:Calculations}.

\subsection{Magnetic modes}

In solving the system of Boltzmann and Einstein equations, one needs to set the initial conditions for the scalar, vector and tensor modes. In Boltzmann codes, such as \textsc{CAMB}, they are set on super-horizon scales, $k\tau \ll 1$, and well after the neutrino decoupling, {\it i.e.} at $\tau \gg \tau_\nu$. After neutrinos decouple from photons, which happens at energies below 1 MeV, they free stream and can develop a non-zero anisotropic stress that compensates the anisotropic stress of the PMF. However, at $\tau < \tau_\nu,$ neutrinos are bound in a tightly coupled fluid with photons and baryons and are unable to compensate for the magnetic anisotropic stress, which then acts as a source of adiabatic scalar and tensor mode perturbations \cite{Lewis:2004ef}. The latter are usually assumed to be uncorrelated with the adiabatic fluctuations generated by inflation and are treated as separate \emph{passive} magnetic modes \cite{Lewis:2004ef,Shaw:2010}. After the neutrino decoupling, PMFs generate isocurvature type perturbations, in which the neutrino anisotropic stress compensates the anisotropic stress of the PMF, known as the \emph{compensated modes} \cite{Giovannini:2004aw,Giovannini:2006kc,Finelli:2008xh}. It was also realized that, for PMFs generated during inflation, there is another adiabatic mode known as the magnetic inflationary mode \cite{Bonvin:2013tba}. However, its amplitude is strongly model-dependent and, in order to keep our approach as model-independent as possible, we restrict our analysis to PMFs generated after inflation, and consider the passive and compensated modes only. 

The amplitude of the adiabatic scalar mode is set by the comoving curvature perturbation $\zeta$ which, in the absence of a PMF,  is conserved on super-Hubble scales. However, as mentioned above, a PMF present before neutrino decoupling would source the growth of $\zeta$ \cite{Lewis:2004ef}. When neutrinos decouple, their anisotropic stress rapidly grows to compensate the anisotropic stress of the PMF. When the compensation is effective, $\zeta$ stops growing, having reached the final value \cite{Lewis:2004ef,Shaw:2010}
\begin{equation}
\zeta = \zeta(\tau_B) -\frac{1}{3} R_{\gamma} \Pi_B^{(0)} \biggl[ \ln(\tau_{\nu} / \tau_B) + \biggl(\frac{5}{8 R_{\nu}} -1 \biggr) \biggr],
\end{equation}
where $\zeta(\tau_B)$ is the comoving curvature perturbation at the time $\tau_B$, after inflation, at which the PMF was generated. What is commonly referred to as the scalar passive mode, is the part of the adiabatic mode associated with the growth of $\zeta$, proportional to $ \ln(\tau_{\nu} / \tau_B)$, and its power spectrum is proportional to $\braket{\Pi_B^{(0)  *}\, \Pi_B^{(0)}}$ \cite{Lewis:2004ef,Shaw:2010}. 

The compensated scalar modes start being actively sourced by the PMF after neutrino-decoupling. There are two such modes, sourced by $\Delta_B$ and $\Pi_B^{(0)}$, with power spectra proportional to $\braket{\Delta_B^* \Delta_B}$ and $\braket{\Pi_B^{(0) *} \, \Pi_B^{(0)}}$ respectively. The two scalar compensated modes are not independent, as the correlation $\braket{\Delta_B^* \Pi_B^{(0)}}$ does not vanish.

Since vector perturbations rapidly decay if not continuously sourced, there are no passive vector modes. The only regular solution for the vector part of the Einstein-Boltzmann system, with a PMF as an active source, is a compensated mode for which the anisotropic stress $\Pi_B^{(\pm 1)}$ is compensated by $\Pi_{\nu}^{(\pm 1)}$.

Tensor modes, like scalar modes, have both passive and compensated modes. Before neutrino decoupling, $\Pi_B^{(\pm 2)}$ sources the tensor perturbation, generating the tensor passive mode proportional to $\ln (\tau_{\nu}/\tau_B)$. After neutrino decoupling, the anisotropic stress of the PMF is compensated, leading to the tensor compensated mode. The latter are entirely negligible compared to other modes \cite{Lewis:2004ef,Shaw:2010} and can be safely neglected when deriving CMB bounds on the PMF.

\begin{figure}[!htbp]
\centering
\includegraphics[width=0.45\textwidth]{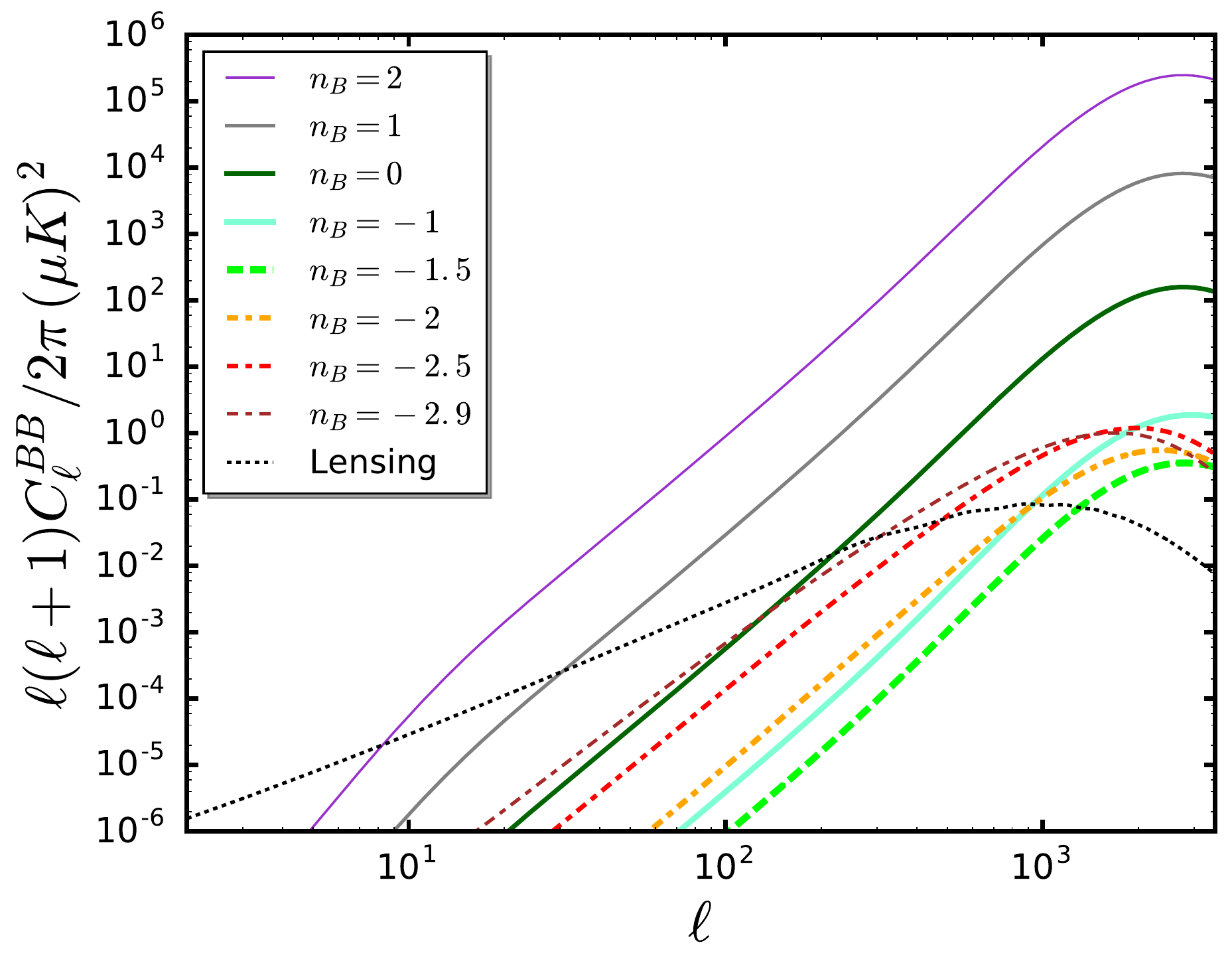}
\caption{\label{fig:VecBmode} The B-mode spectrum from the PMF vector mode with $B_{1 \, \mathrm{Mpc}} = 4.5 \, \mathrm{nG}$ and different values of the spectral index $n_B$. The dotted line shows the lensing contribution.}
\end{figure}

\begin{figure}[!htbp]
\centering
\includegraphics[width=0.45\textwidth]{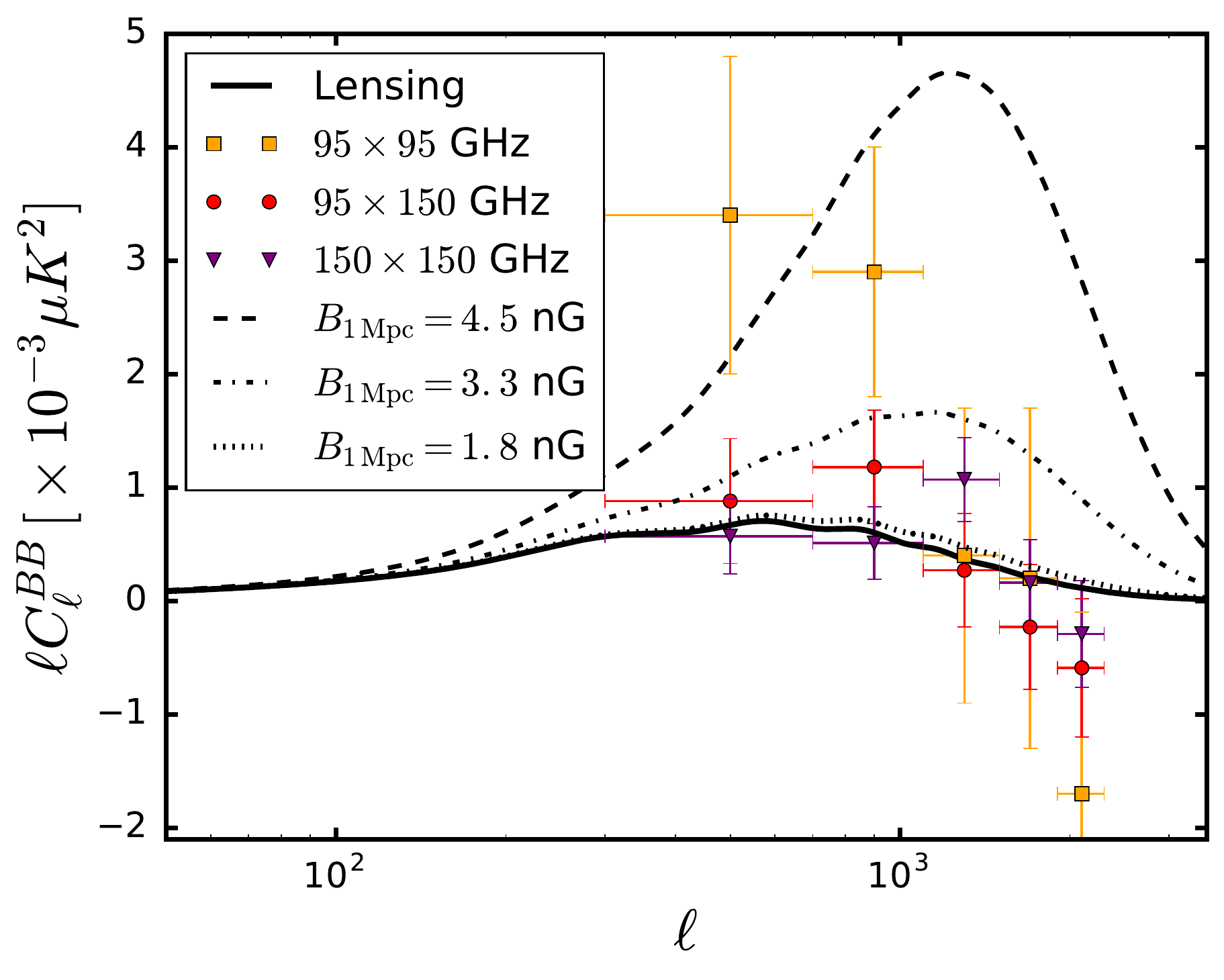}
\caption{\label{fig:BmodeSPT} 
The B-mode spectrum from the PMF vector mode added to the lensing contribution (solid black line) for $B_{1 \, \text{Mpc}}=4.5$ nG (dashed line), $B_{1 \, \text{Mpc}} = 3.3$ nG (dot-dashed line) and $B_{1 \, \text{Mpc}} = 1.8$ nG  (dotted line), with $n_B=-2.9$. The three \textsc{SPT} bandpowers are shown in orange, red and purple.} 
\end{figure}

In Fig.~\ref{fig:MagneticModes} we show contributions of the magnetic modes to the CMB spectra for a PMF with $B_{1 \, \text{Mpc}} = 4.5$ nG, $n_B=-2.9$ and $\beta=\log_{10} (\tau_{\nu}/\tau_B) = 17$, corresponding to the PMF generation energy scale of $10^{14}$ GeV. In addition, in Fig.~\ref{fig:VecBmode}, we show the vector mode contribution to BB for different values of the spectral index, which is representative of the way all CMB spectra vary with $n_B$. Note the qualitative change in the dependence on $n_B$ that occurs at $n_B=-1.5$.  A change in $n_B$ in the $-3<n_B<-1.5$ range alters the shape of the CMB spectra, with an increase in $n_B$ causing a shift of power from lower to higher $\ell$. This reduces the CMB anisotropy power on scales within the observational window. In contrast, for $n_B>-1.5$, CMB spectra become cutoff dominated and scale as white noise, with the shape being practically independent of $n_B$. In that regime, larger values of $n_B$ lead to more CMB power for the same PMF strength on $1$ Mpc scale.

The contribution of the magnetic vector mode to the B-mode polarization power spectrum, can be well constrained by the current and future CMB experiments capable of detecting the B-modes from weak lensing. As illustrated in Fig.~\ref{fig:BmodeSPT}, the large-$\ell$ measurements of the B-mode polarization performed by the \textsc{SPT} can place competitive bounds on the amplitude of the PMF. In fact, as shown later in the paper, they significantly reduce the upper bound obtained from \textsc{Planck}.

\subsection{Magnetic patches for CAMB and CosmoMC}

We developed a patch\footnote{The patch is publicly available at \url{https://alexzucca90.github.io/MagCAMB/}} for the publicly available Boltzmann code \textsc{CAMB} \cite{Lewis:1999bs}. The main features are briefly summarized below, while the details of the implementation will be explained elsewhere \cite{Yun:2016}.

We introduce the effects of the PMF into the Einstein and Boltzmann equations closely following the formalism of \cite{Shaw:2010} and the associated code by Shaw and Lewis (SL). Among the notable improvements with respect to SL is making the code compatible with \textsc{CosmoMC} and extending the allowed range of the magnetic spectral index to values $n_B \ge -1.5$. We recalculated the integrals involved in the correlation functions of the magnetic perturbations $\Delta_B$ and $\Pi_B^{(0,1,2)}$. The upper integration bound of the integrals is the ratio $k_D/k$. We confirmed that, for $-3 < n_B <-1.5$, the integrals depend weakly on $k_D/k$. Since the $k$ modes involved in the computation of the CMB power spectra are much smaller than the damping scale $k_D$, we computed the integrals in the approximate limit $k_D/k \to \infty$. For arbitrary $n_B$ in that range, we interpolate on a grid of precomputed integrals. For $n_B \ge -1.5$, the integrals depend strongly on the upper integration bound $k_D/k$. Since for arbitrary $n_B$ and $k$ the integrals involve the hypergeometric function, we sampled the integrals and computed a set of fitting functions for each correlation function as in \cite{Paoletti:2010}.

We also have extended the latest version of the \textsc{CosmoMC} code \cite{Lewis:2002ah} to include the contributions of the scalar, vector and tensor compensated and passive magnetic modes\footnote{The patch is publicly available at \url{https://github.com/alexzucca90/MagCosmoMC}}. 

\section{Bounds on PMF from current CMB data}
\label{sect:DataAnalysis}
We use the measurements of the CMB anisotropies power spectra by the \textsc{Planck} satellite \cite{Aghanim:2015xee} and the measurements of the CMB B-mode polarization by the \textsc{SPT} \cite{Keisler:2015hfa} to constrain the amplitude, the spectral index and the time of generation of the primordial magnetic fields. 

For \textsc{Planck}, we use the joint TT, TE, EE and BB likelihood in the range $2<\ell<29$, denoted as \textsc{LowTEB}, together with the high-$\ell$ temperature likelihood in the range $30<\ell<2508$, simply denoted as \textsc{TT}. We also consider the case in which the TT likelihood is replaced with the joint TT, TE and EE polarization likelihood (denoted as \textsc{TTTEEE}). We also perform the analysis with and without using the likelihood from the \textsc{BICEP2/Keck-Planck} (BKP) cross correlation analysis \cite{Ade:2015tva}.

The SPT likelihood \cite{Keisler:2015hfa, SPTlikel} is a multivariate Gaussian likelihood and uses three bandpowers from the $95 \text{GHz} \times 95 \text{GHz}$, $95 \text{GHz} \times 150 \text{GHz}$ and $150 \text{GHz} \times 150 \text{GHz}$ spectra. It also takes into account the contributions to B-modes from the dust emission within our Galaxy and from the polarized emission from extragalactic sources. The dust emission is modelled according to Eq.~(21) in \cite{Keisler:2015hfa} and is scaled by an overall dust emission amplitude $A_{\text{dust}}$ \cite{Adam:2014bub}. The extragalactic sources are modelled through a constant $C_l $ term with different amplitudes for each bandpower, $A_{\text{PS}, 95}$, $A_{\text{PS}, 95 \times 150}$ and $A_{\text{PS}, 150}$. These nuisance parameters are marginalized over using priors shown in Tabel~\ref{tab:PriorsDustNuisParam}. We have extended the \textsc{SPT} likelihood code to include the contributions of the PMF to the CMB B-modes.

\begin{table}[!htbp]
\centering
\begin{tabular}{ l | c | c }
 Parameter & Prior & \\
 \hline \hline
 $A_{\text{dust}}$ \dotfill & $[0.0,  2.5]$  & Gaussian \\  
 $A_{\text{PS}, 95}$ \dotfill & $[0.0, 4.0]$ & flat \\
 $A_{\text{PS}, 95 \times 150}$ \dotfill & $[0.0, 4.0]$ & flat \\
 $A_{\text{PS}, 150}$ \dotfill & $[0.0 4.0]$ & flat 
\end{tabular}
\caption{\label{tab:PriorsDustNuisParam} Priors on the nuisance parameters used in the \textsc{SPT} likelihood described in Sect.~\ref{sect:PlanckSPTDataAnalysis}.}
\end{table}

We assume a flat universe and, as in \cite{Ade:2015cva}, restrict our analysis to three massless neutrinos. We also assume that the primary (inflationary) and the passive and compensated magnetic modes are uncorrelated, so that their contributions to the CMB spectra can be calculated separately and simply added as
\begin{equation}
\label{eqn:sumCls}
C^{\text{theor}}_{\ell} = C_{\ell}^{\text{prim}} + C_{\ell}^{\text{pass}} + C_{\ell}^{\text{comp}}.
\end{equation}
A scenario with correlated inflationary and magnetic modes has been discussed in \cite{Kunze:2013hy} in a context of a specific model. We account for the effect of weak lensing by large scale structure on the primary mode only, and we marginalize over astrophysical residuals \cite{Aghanim:2015xee, Ade:2015cva}. 

The pivot Fourier number for the primary primordial spectrum is set to $k_{*} = 0.05 \text{Mpc}^{-1}$, while the magnetic smoothing scale is set to $\lambda =1$Mpc. We vary the baryon density $\omega_b = \Omega_b h^2$, the CDM density $\omega_c = \Omega_c h^2$, the reionization optical depth $\tau_{\text{reion}}$, the ratio of the sound horizon to the angular diameter distance at decoupling $\theta$, and the amplitude $A_s$ and the spectral index $n_s$ of the primary primordial spectrum of curvature perturbations. We also vary the additional magnetic parameters $B_{1 \text{Mpc}}$, $n_B$ and $\beta = \log_{10}(\tau_{\nu} / \tau_B)$ . The priors assumed on the parameters are given in Table~\ref{tab:PriorsCosmoParam}.

\begin{table}[!htbp]
\centering
\begin{tabular}{ l | c }
 Parameter & Flat Prior \\
 \hline \hline
$\omega_b$ \dotfill & $[0.005,0.1]$   \\  
$\omega_c$  \dotfill & $[0.001, 0.99]$  \\
$\tau_{\text{reion}}$ \dotfill & $[0.01, 0.8]$ \\
 $\theta$ \dotfill  & $[0.5,10]$  \\
 $\ln (10^{10} A_s)$ \dotfill & $[2,4]$  \\
 $n_s$  \dotfill & $[0.8,1.2]$  \\
 $r$ \dotfill & $[0,2]$ \\
 \hline
 $B_{1 \text{Mpc}}$ \dotfill & $[0,10]$ \\
 $ \log_{10} (B_{1 \text{Mpc}}/ {\rm nG})$ \dotfill & $[-5,1]$ \\
 $\beta=\log_{10} (\tau_{\nu}/ \tau_B)$ \dotfill & $[4,17]$  \\
 $n_B$ \dotfill & $[-2.9,3]$  
\end{tabular}
\caption{\label{tab:PriorsCosmoParam} Priors on the parameters varied in the MCMC analysis. We performed the analysis separately with the uniform and logarithmic priors on $B_{1 \text{Mpc}}$.}
\end{table}

As can be seen from Fig.~\ref{fig:MagneticModes}, for nearly scale-invariant PMFs, the passive tensor magnetic mode is similar in shape to the primary (inflationary) tensor mode, with an amplitude that depends on the time of the generation of the PMF, $\beta=\log_{10}(\tau_{\nu}/\tau_B)$. To address a potential degeneracy between the tensor-to-scalar ratio $r = A_T/ A_s$ and $\beta$, we consider the cases with a fixed $r=0$, as well as with co-varying the two parameters. 

\subsection{Constraints from Planck data}
\label{sect:PlanckDataAnalysis}

\begin{figure}[!htbp]
\centering
\includegraphics[width=0.45\textwidth]{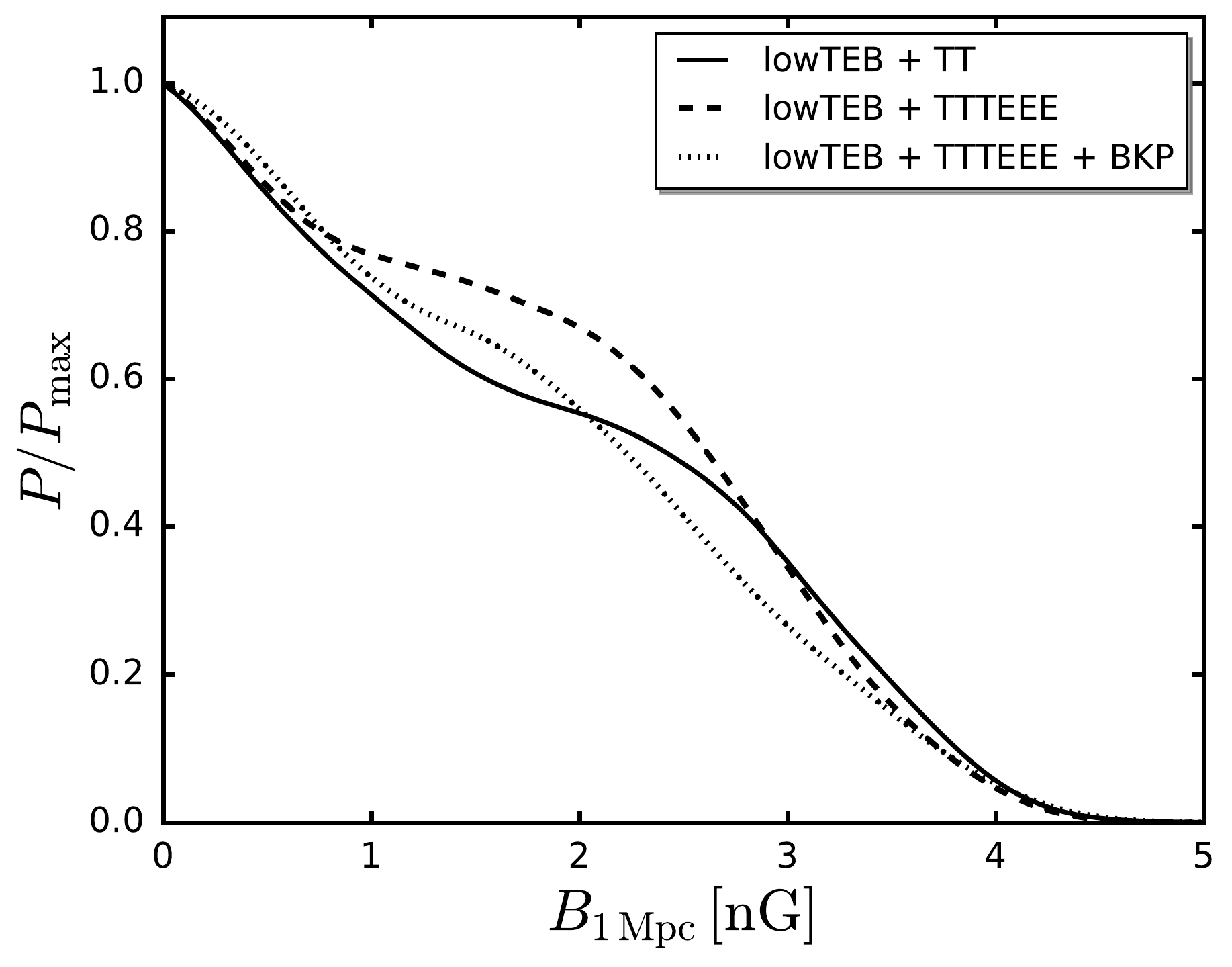} 
\caption{\label{fig:PlanckPosterior} The probability distribution function for the magnetic amplitude $B_{1 \, \text{Mpc}}$ from the \textsc{Planck} data sets described in Sect.~\ref{sect:PlanckDataAnalysis}. We show only the case with $r=0$ since varying $r$ does not affect the results.}
\end{figure}

To derive constraints on the PMF from \textsc{Planck}, we use the \textsc{Planck} likelihood code described in detail in \cite{Aghanim:2015xee}.
A thorough analysis has already been conducted by the \textsc{Planck} collaboration in \cite{Ade:2015cva}. Since scalar passive modes are not supposed to contribute significantly to the magnetic signals in the CMB (as shown in Sec.~\ref{sect:PMFImpact}), the authors of \cite{Ade:2015cva} included them only in the special case of a nearly scale invariant PMF with $n_B = -2.9$. Conversely, we include scalar passive modes in all of our analysis for the sake of completeness.


Fig.~\ref{fig:PlanckPosterior} shows the marginalized probability distribution function (PDF) for $B_{1 \, \text{Mpc}}$ derived from \textsc{Planck} data. The Figure only shows the case with $r=0$, since the PDF in the case of co-varied $r$ was essentially the same. The 95\% CL bounds on $B_{1 \, \text{Mpc}}$ are summarized in Tab.~\ref{tab:PlanckConstraints}, including the case with co-varied $r$.

\begin{table}[!htbp]
\centering
\begin{tabular}{l c}
Data sets & $B_{1 \, \text{Mpc}} / \text{nG}$ \\
\hline \hline
\textsc{LowTEB + TT}, $r = 0$ \dotfill & $<3.3$ \\
\textsc{LowTEB + TT}, $r$ free \dotfill & $<3.3$ \\
\textsc{LowTEB + TTTEEE}, $r = 0$ \dotfill & $<3.2$ \\
\textsc{LowTEB + TTTEEE}, $r$ free \dotfill & $<3.2$ \\
\textsc{LowTEB + TTTEEE + BKP}, $r = 0$ \dotfill & $<3.3$ \\
\textsc{LowTEB + TTTEEE + BKP}, $r$ free \dotfill & $<3.3$ 
\end{tabular}
\caption{\label{tab:PlanckConstraints} Upper bounds (95\% CL) for the PMF amplitude $B_{1 \, \text{Mpc}}$ obtained from the combination of \textsc{Planck} data sets described in Sect.~\ref{sect:PlanckDataAnalysis}. The magnetic spectral index $n_B$ and the PMF generation epoch parameter $\beta$ are unconstrained.} 
\end{table}

The magnetic spectral index $n_B$ and the PMF generation epoch parameter $\beta$ are unconstrained. We discuss these parameters in more detail in the next subsection.

\subsection{Constraints from Planck combined with \textsc{SPT}}
\label{sect:PlanckSPTDataAnalysis}

\begin{figure*}[!htbp]
\centering
\subfigure{
\includegraphics[width=0.45\textwidth]{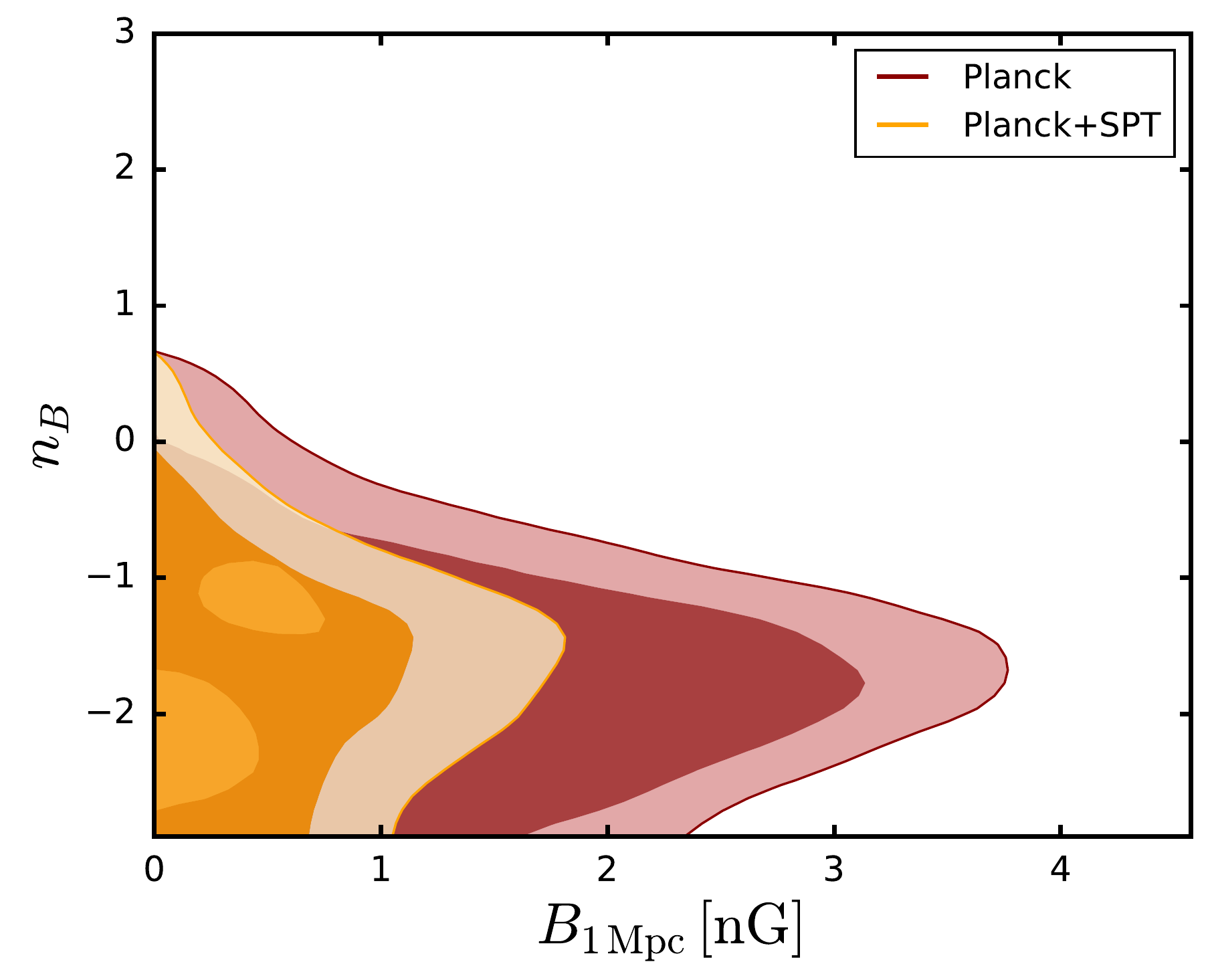}} \,
\subfigure{
\includegraphics[width=0.45\textwidth]{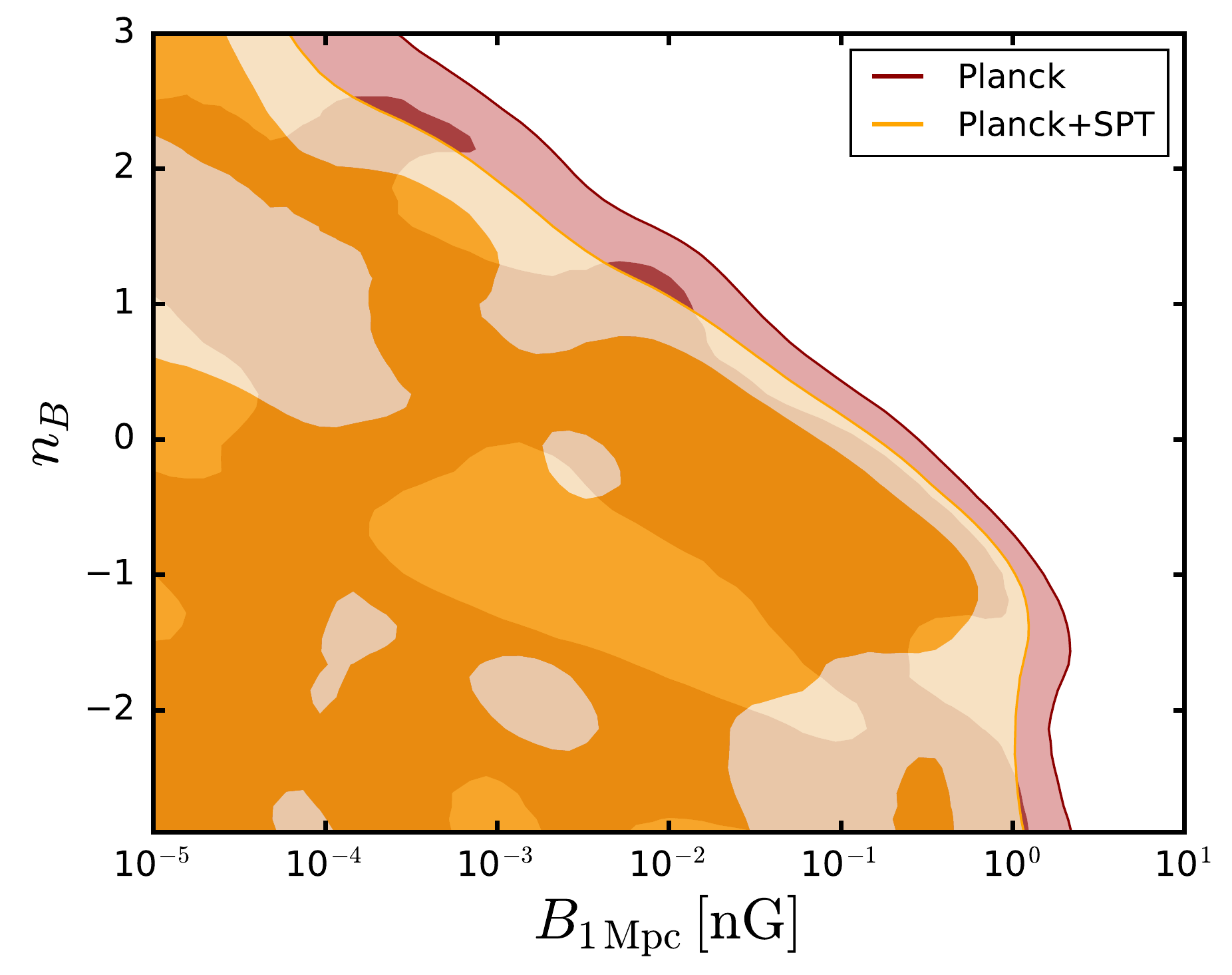}}
\caption{\label{fig:PlanckContours} Left panel: the joint probability for the magnetic amplitude $B_{1 \, \text{Mpc}}$ and the magnetic index $n_B$ using uniform prior on $B_{1 \, \text{Mpc}}$.  Right panel: the joint probability for  $B_{1 \, \text{Mpc}}$ and  $n_B$ using uniform prior on $\log_{10}(B_{1 \, \text{Mpc}}/\mathrm{nG})$.  The two shaded regions represent the 68\%C.L. and 95\% C.L. respectively.  The apparent bound on $n_B$ in the left panel disappears when using the logarithmic prior, as shown in the right panel.}
\end{figure*}

\begin{figure}[!htbp]
\centering
\includegraphics[width=0.45\textwidth]{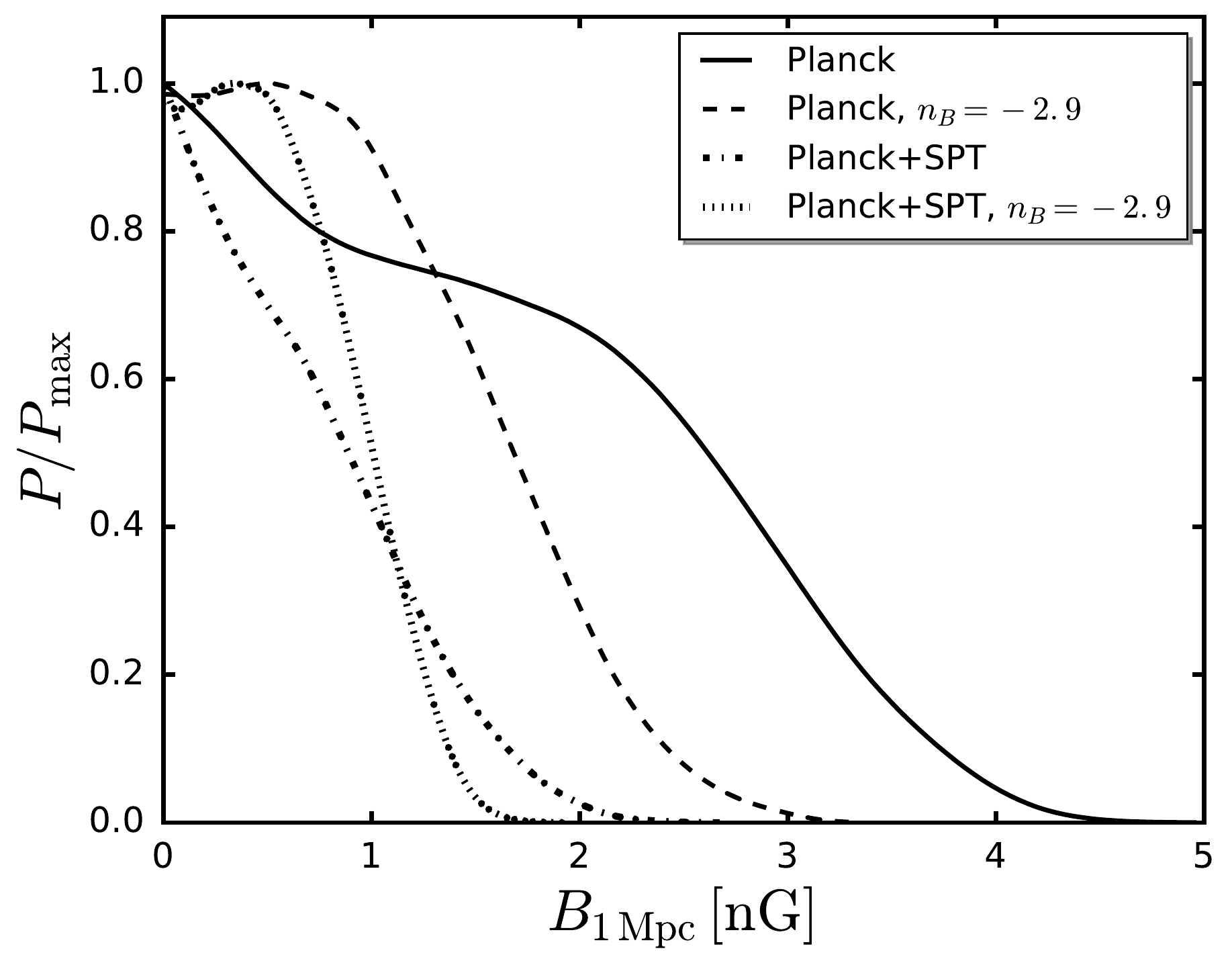}
\caption{\label{fig:SPTPosterior} The marginalized PDFs for the magnetic amplitude $B_{1 \, \text{Mpc}}$ from  \textsc{Planck} and the combination of \textsc{Planck} and \textsc{SPT}. We only show the PDFs obtained with $r=0$, as the case with co-varying $r$ is essentially the same. We also show the PDFs for the nearly scale-invariant case, $n_B=-2.9$. }
\end{figure}

Combining \textsc{Planck} with the B-mode polarization spectrum from \textsc{SPT} significantly tightens the bounds on the PMF, because of the contribution of the PMF vector modes, as illustrated in Fig.~\ref{fig:BmodeSPT}.  We perform the analysis using the \textsc{SPT} likelihood and the \textsc{Planck} \textsc{lowTEB} and \textsc{TTTEEE} likelihoods, referring to the combination of them as \textsc{Planck} for simplicity. We do not include the BKP data, after the analysis in the previous subsection confirmed that it does not affect the bounds on the PMF.

In Fig.~\ref{fig:PlanckContours}, we show the joint probability for the magnetic amplitude $B_{1 \, \text{Mpc}}$ and the magnetic index $n_B$ from Planck alone and after combining \textsc{Planck} with \textsc{SPT}. The two parameters are correlated, with the bound on $B_{1 \, \text{Mpc}}$ becoming weaker with increasing $n_B$ in the $-3<n_B<-1.5$ range, and stronger for $n_B>-1.5$. This is due the qualitative change in the dependence of the CMB spectra on the magnetic power spectrum that occurs at $n_B=-1.5$.  Namely, as illustrated in Fig.~\ref{fig:MagneticModes}, an increase in $n_B$ results in a shift of power from lower to higher $\ell$, reducing the CMB power on scales inside the observational window and thus allowing for larger PMF strengths. In contrast, for $n_B>-1.5$, the shapes of the CMB spectra are cutoff dominated, with larger $n_B$ resulting in more CMB power for the same PMF strength on $1$ Mpc scale, leading to tighter constraints on $B_{1 \, \text{Mpc}}$.

Fig.~\ref{fig:PlanckContours} separately shows the cases with a uniform (left panel) and the logarithmic (right panel) priors on $B_{1 \, \text{Mpc}}$. As expected, the apparent upper bound on $n_B$, present in the case of the uniform prior and also observed in \cite{Ade:2015cva}, is not physical and disappears in the case of the logarithmic prior. Indeed, there cannot be a bound on the spectral index of the PMF spectrum without a positive detection of the amplitude. The PDFs for the amplitude $B_{1 \, \text{Mpc}}$, after marginalizing over $n_B$, are shown in Fig.~\ref{fig:SPTPosterior}.

Two values of $n_B$ are of particular theoretical interest. The first and simplest models of inflationary magnetogenesis \cite{Turner:1987bw,Ratra:1991bn} predict a nearly scale-invariant PMF with $n_B \approx -3$. The combined bound from \textsc{Planck} and \textsc{SPT} on the nearly scale-invariant PMF ($n_B=-2.9$)\footnote{To avoid divergent integrals, we restrict our analysis to $n_B \ge -2.9$. We also note that the dependence on the smoothing scale disappears and $B_{1 \, \text{Mpc}} = B_{\rm eff}$ for scale-invariant fields.} is $B_{1 \, \text{Mpc}} \approx B_{\rm eff} < 1.2$ nG at 95\% CL. The corresponding bound from \textsc{Planck} alone is $2.0$ nG, in agreement with \cite{Ade:2015cva}. 

The PMFs generated in post-inflationary phase transitions have small coherence lengths and are uncorrelated on cosmological scales. Causality forces the spectra of such fields to have $n_B = 2$ on scales of relevance to CMB anisotropies \cite{Jedamzik:1996wp,Durrer:2003ja,Jedamzik:2010cy}. For such fields, we find $B_{1 \, \text{Mpc}}< 0.002$ nG at 95\% CL. However, since most of the power of the causally generated PMFs is concentrated near the cutoff scale $2\pi /k_D \ll {\rm 1 Mpc}$, using $B_{1 \, \text{Mpc}}$ to quantify their amplitude can be misleading. Instead, it is more appropriate to use $\Omega_{B\gamma}$ or $B_{\rm eff}$, which are representative of the total PMF energy density \cite{Kahniashvili:2009qi,Kahniashvili:2010wm,Pogosian:2011qv}. Using the conversion in Eq.~(\ref{eq:beff}), we derive  $\Omega_{B\gamma} < 10^{-3}$ or $B_{\rm eff} < 100$ nG at 95\% CL. For reference, the Big Bang Nucleosynthesis constrains the magnetic fraction to be $\Omega_{B\gamma} \lesssim 0.1$ \cite{Kernan:1995bz,Grasso:1996kk,Cheng:1996yi}.

\begin{table}[!htbp]
\centering
\begin{tabular}{c|ccc}
& $B_{1 \, \text{Mpc}}/{\rm nG}$ & $B_{\rm eff}/{\rm nG}$ & $\Omega_{B\gamma}$  \\ 
\hline \hline
$n_B$ marginalized  & $<1.5$ & n/a & n/a  \\
$n_B =-2.9$ & $<1.2$  & $<1.2$  & $<1.4\times 10^{-7}$ \\
$n_B = 2$ & $<0.002$  & $<100$  & $<10^{-3}$
\end{tabular}
\caption{\label{tab:PlanckSPTConstraints} Upper bounds (95\% CL) on the PMF amplitude $B_{1 \, \text{Mpc}}$, the effective PMF strength $B_{\rm eff}$ and the magnetic density fraction $\Omega_{B\gamma}$ obtained from \textsc{Planck} and \textsc{SPT}.} 
\end{table}

Constraints on the PMF strength after marginalizing over $n_B$, as well as for the two special cases of theoretical interest, are summarized in Table~\ref{tab:PlanckSPTConstraints}.  

\begin{figure}[!htbp]
\includegraphics[width = 0.45 \textwidth]{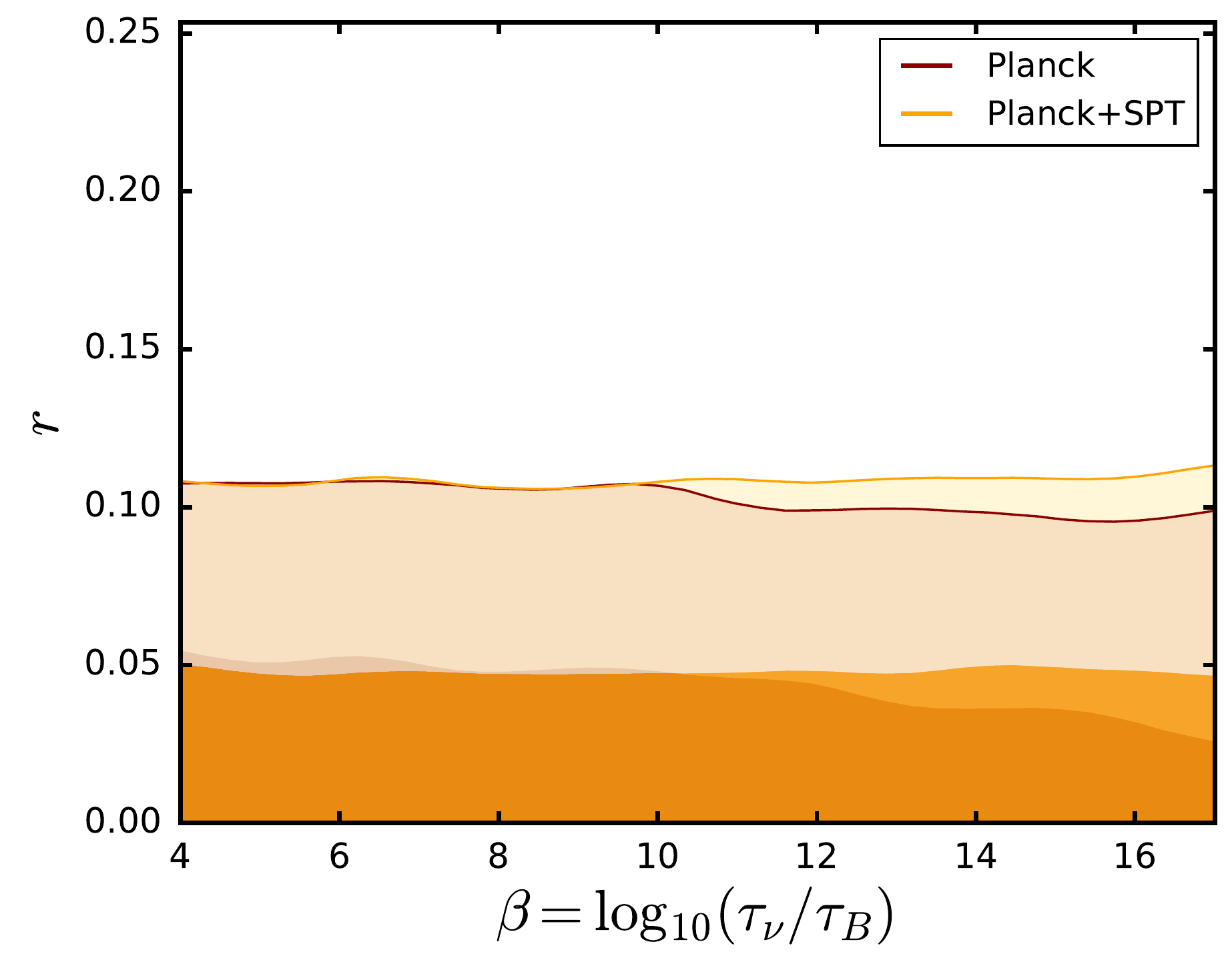}
\caption{\label{fig:r-beta} The joint probability for the scalar to tensor ratio $r$ and the time of generation of the PMF $\log_{10}(\tau_{\nu}/\tau_B)$.  The two shaded regions represent the 68\%CL and 95\% CL, respectively.  }
\end{figure}

The joint probabilities for $r$ and $\beta=\log_{10}(\tau_{\nu}/ \tau_B)$, after marginalizing over other parameters, are shown in Fig.~\ref{fig:r-beta}. It is evident that there is no degeneracy between them and that the time of the generation of the PMF is not constrained by data. This is because the contribution of the passive scalar and tensor modes to TT, TE and EE are too small even for the maximum allowed value of $\beta=17$. As one can see from Fig.~\ref{fig:MagneticModes}, the passive tensor mode is comparable in amplitude to the primary TT at low $\ell$ when $B_{1 \, \text{Mpc}} =4.5$ nG and $\beta =17$.  For smaller values of $\beta$, even higher PMF strengths would be required for the passive tensor mode to be relevant. Such high values of PMF are not allowed because of the PMF vector mode contribution to TT at high $\ell$. Any remaining weak sensitivity to $\beta$ is further diluted by degeneracies with $A_s$, $n_s$ and $\tau_{\rm reion}$. Thus, the passive tensor mode contribution  to TT at low $\ell$ is irrelevant for the strengths of PMF allowed by TT at high $\ell$. The scalar passive mode is even less irrelevant, as evident from Fig.~\ref{fig:MagneticModes}. Note that adding the BPK B-mode data does not make a big difference because of large uncertainties at $\ell$ where the contribution from the passive tensor mode is prominent. Adding the \textsc{SPT} data does not help in constraining $\beta$ either, because \textsc{SPT} only constrains the vector mode contribution to BB and does not add information on scales relevant to the passive tensor mode.

The role of the astrophysical residuals included in the \textsc{Planck} and \textsc{SPT} likelihoods is discussed in Appendix~\ref{sect:AstroPhysicalResiduals}.

\section{Summary and Discussion}
\label{sect:Discussion}

We have derived the bound on the strength of the primordial magnetic field from the \textsc{SPT} CMB B-mode polarization measurements in combination with the CMB temperature and polarization data  from \textsc{Planck}. Adding the \textsc{SPT} information significantly tightens the bound, as it constrains the PMF vector mode contribution to B-modes at $\ell \sim 1000$. Specifically, adding the \textsc{SPT} data reduces the 95\% CL bound on $B_{1 \, \text{Mpc}}$, after marginalizing over the magnetic spectral index $n_B$, from $3.3$ nG to $1.5$ nG. For a nearly scale-invariant PMF with $n_B=-2.9$, the bound is reduced from $2$ nG to $1.2$ nG. The effective strength of a PMF generated in post-inflationary phase transitions, with $n_B=2$, is constrained to $B_{\rm eff} < 100$ nG, corresponding to $\Omega_{B\gamma} < 10^{-3}$, or $B_{1 \, \text{Mpc}}<0.002$ nG, at 95\% CL.

Our results, as well as those obtained by the POLARBEAR collaboration in \cite{Ade:2015cao}, demonstrate that one can extract competitive information about PMFs even from the existing B-mode measurements. Future CMB experiments, in addition to significantly improving the measurement of the B-mode spectrum at high $\ell$, will eventually provide reliable data on scales relevant for the inflationary tensor mode. Such data will help to constrain the passive tensor PMF mode and, thus, the time of the generation of the PMF. Future CMB experiments will also constrain the parity-odd TB and EB correlations, leading to meaningful bounds on the helical component of the PMF, which has been neglected in the present work. They will also tightly constrain the mode-coupling correlations induced by the Faraday rotation of CMB polarizations. The Faraday rotation angle is linear in $B_{1 \, \text{Mpc}}$, while CMB anisotropies scale as a square of the PMF strength (so that the CMB spectra scale as  $B^4_{1 \, \text{Mpc}}$), and, with the improved sensitivity and resolution of upcoming experiments, can reduce the upper bound on $B_{1 \, \text{Mpc}}$ by an order of magnitude \cite{Yadav:2012uz,De:2013dra,Pogosian:2013dya}.

The pioneering work by the POLARBEAR \cite{Ade:2014afa}, \textsc{BICEP/Keck} \cite{Ade:2014xna,Array:2015xqh} and \textsc{SPT} \cite{Keisler:2015hfa} collaborations has ushered cosmology into the era of precision B-mode science. In addition to searching for signatures of inflationary gravitational waves and primordial magnetic fields, B-modes will be used to probe the neutrino masses \cite{Abazajian:2013oma}, modifications of gravity \cite{Amendola:2014wma,Raveri:2014eea}, cosmic (super)strings \cite{Moss:2014cra,Avgoustidis:2011ax} and other fundamental physics \cite{Abazajian:2016yjj}.

\begin{acknowledgments}
We thank Richard Shaw for sharing with us the updated version of his magnetic \textsc{CAMB} patch. We benefited from communications and discussion with Camille Bonvin, Fabio Finelli, Alireza Hojjati, Antony Lewis, Tina Kahniashvili, Daniela Paoletti, Richard Shaw and Tanmay Vachaspati. We acknowledge support by the Natural Sciences and Engineering Research Council of Canada (NSERC). This research was enabled in part by support provided by WestGrid \cite{westgrid} and Compute Canada \cite{compute}. AZ is supported in part by the Bert Henry Memorial Entrance Scholarship at SFU.

\end{acknowledgments}

\appendix
\section{Correlation Functions of the Magnetic Energy Momentum Tensor }
\label{sect:Calculations}

The Fourier transform of $T_{B \, j}^i$ in equation \eqref{eqn:PMFEnergyMomentumTensor} is given by
\begin{equation}
\begin{split}
T_{B  j}^{\,i} (\mathbf{k}) = & \frac{1}{4 \pi a^4 (2 \pi)^3} \times \\ &
\int d^3q \, \biggl[  \frac{1}{2} \delta^i_j B_l(\mathbf{q}) B^l (\mathbf{k}-\mathbf{q}) -  B^i(\mathbf{k}) B_j (\mathbf{k}-\mathbf{q})\biggr].
\end{split}
\end{equation}
Using the previous equation, the magnetic perturbations $\Delta_B$ and $\Pi^i_{B\,j}$ are given by
\begin{gather}
\Delta_B (\mathbf{k}) = \frac{1}{3 \, p_{\gamma}} T_{B \,i}^{\,i} = \frac{1}{8 \pi \rho_{\gamma}^0 (2 \pi)^3} \int d^3q \, B_l(\mathbf{q}) B^l(\mathbf{k} - \mathbf{q} ), \\
\begin{split}
\Pi_{B j}^{\, i} (\mathbf{k}) = & \frac{1}{p_{\gamma}}\biggl( T_{B  j}^{\,i} - \frac{1}{3} \delta^i_j \, T_{B  n}^{\, n} \biggr) = \frac{3}{4 \pi \rho_{\gamma}^0 (2 \pi)^3} \times \\
 & \int d^3q \,\biggl[  \frac{1}{3} \delta^i_j B_l(\mathbf{q}) B^l(\mathbf{k}-\mathbf{q}) - B^i(\mathbf{q}) B_j(\mathbf{k} - \mathbf{q}) \biggr].
\end{split}
\end{gather}
The scalar, vector and tensor components of $\Pi_{B j}^{\,i}$ are obtained by the scalar products $\Pi^{ij}_B Q_{ij}^{(0, \pm 1,\pm 2 )}$, and are respectively
\begin{gather}
\begin{split}
\Pi_B^{(0)} (\mathbf{k}) & =   \frac{3}{2} Q_{ij}^{(0)}(\mathbf{k}) \Pi_B^{ij}(\mathbf{k}) = - \frac{3}{2} \frac{3}{4 \pi \rho_{\gamma}^0 (2 \pi)^3} \times \\
&\int d^3q \, \biggl[ \frac{1}{3}B_l(\mathbf{q})B^l(\mathbf{k} - \mathbf{q}) -\hat{k}_i B^i(\mathbf{q}) \hat{k}_j B^j(\mathbf{k}-\mathbf{q})\biggr], 
\end{split} \\ 
\begin{split}
\Pi_B^{(\pm 1)} (\mathbf{k}) & =2  Q_{ij}^{(\mp 1)}(\mathbf{k}) \Pi_B^{ij}(\mathbf{k}) = - \frac{3 i}{4 \pi \rho_{\gamma}^0 (2 \pi)^3}  \int d^3q \times \\
& \biggl[ \hat{k}_i B^i(\mathbf{q}) e^{(\mp)}_jB^j(\mathbf{k} - \mathbf{q}) + \hat{k}_i B^i(\mathbf{k} - \mathbf{q}) e^{(\mp)}_jB^j(\mathbf{q}) \biggr],
\end{split} \\
\begin{split}
\Pi_B^{(\pm 2)}(\mathbf{k}) & = \frac{2}{3} Q_{ij}^{(\mp 2)}(\mathbf{k})\Pi_B^{ij}(\mathbf{k}) = - \sqrt{\frac{2}{3}} \frac{3}{4 \pi \rho_{\gamma}^0 (2 \pi)^3} \times \\
& \int d^3q \, e_i^{(\mp)}B^i(\mathbf{q}) e_j^{(\mp)}B^j(\mathbf{k}- \mathbf{q}).
\end{split}
\end{gather}
We then use the equations above to compute the correlation functions of the magnetic perturbations. We define $\beta = \hat{\mathbf{k}} \cdot \widehat{(\mathbf{k-q})}$, $\gamma = \hat{\mathbf{k}} \cdot \hat{\mathbf{q}}$ and $\mu = \hat{\mathbf{q}} \cdot \widehat{(\mathbf{k-q})} $. In the scalar sector we have three correlation functions,
\begin{equation}
\begin{split}
& \braket{\Delta_B(\mathbf{k})\Delta_B^*(\mathbf{k}^{\prime})}  \\
& = \frac{\delta^{(3)}(\mathbf{k}-\mathbf{k}^{\prime})}{32 \pi^2 (\rho_{\gamma}^0)^2} \int d^3q \, (1+\mu^2)P_B(q) P_B (|\mathbf{k} - \mathbf{q}|),
\end{split}
\end{equation}
\begin{equation}
\begin{split}
& \braket{\Delta_B(\mathbf{k}) \Pi_B^{(0) *}(\mathbf{k}^{\prime})}  \\
& = \frac{3 \, \delta^{(3)}(\mathbf{k}- \mathbf{k}^{\prime})}{16 \pi^2 (\rho_{\gamma}^0)^2} \int d^3q \, P_B(q) P_B(|\mathbf{k}-\mathbf{q}|)  \\
& \times \biggl[ 1- \frac{1}{2} \mu^2 - \frac{3}{2}(\beta^2 + \gamma^2) + \frac{3}{2} \mu \gamma \beta \biggr],
\end{split}
\end{equation}
\begin{equation}
\begin{split}
& \braket{\Pi_B^{(0)}(\mathbf{k}) \Pi_B^{(0)*}(\mathbf{k}^{\prime})}   \\
&= \frac{9\,  \delta^{(3)}(\mathbf{k} - \mathbf{k}^{\prime})}{8 \pi^2 (\rho_{\gamma}^0)^2} \int d^3q \, P_B(q) P_B(|\mathbf{k} - \mathbf{q}|) \\
& \times \biggl[ 1 + \frac{1}{4} \mu^2 - \frac{3}{4} (\gamma^2 + \beta^2) - \frac{3}{2} \mu \gamma \beta + \frac{9}{4} \gamma^2 \beta^2 \biggr].
\end{split}
\end{equation}

The vector two-point correlation function is 
\begin{equation}
\begin{split}
& \braket{\Pi_B^{(1)}(\mathbf{k}) \Pi_B^{(1)*}(\mathbf{k}^{\prime})} \\
& =  \braket{\Pi_B^{(+1)}(\mathbf{k}) \Pi_B^{(+1)*}(\mathbf{k}^{\prime}) +\Pi_B^{(-1)}(\mathbf{k}) \Pi_B^{(-1)*}(\mathbf{k}^{\prime}) } \\
& = \frac{9 \, \delta^{(3)}(\mathbf{k}-\mathbf{k}^{\prime})}{4 \pi^2 (\rho_{\gamma}^0)^2} \int d^3q \, P_B(q) P_B(|\mathbf{k-q}|) \\
& \times  \biggl[ 1- 2\gamma^2 \beta^2  + \gamma \beta \mu\biggr],
\end{split}
\end{equation}
and, finally, the tensor two-point correlation function is
\begin{equation}
\begin{split}
&\braket{\Pi_B^{(2)}(\mathbf{k}) \Pi_B^{(2)*}(\mathbf{k}^{\prime})} \\
& = \braket{\Pi_B^{(+2)}(\mathbf{k}) \Pi_B^{(+2)*}(\mathbf{k}^{\prime}) +\Pi_B^{(-2)}(\mathbf{k}) \Pi_B^{(-2)*}(\mathbf{k}^{\prime}) } \\
& = \frac{3 \, \delta^{(3)}(\mathbf{k-k^{\prime}})}{8 \pi^2 (\rho_{\gamma}^0)^2} \int d^3q \, P_B(q) P_B(|\mathbf{k-q}|)  (1+\beta^2)(1+\gamma^2).
\end{split}
\end{equation}

\section{Astrophysical Residuals}
\label{sect:AstroPhysicalResiduals}

\begin{figure}[!htbp]
\centering
\subfigure{\includegraphics[width=0.225\textwidth]{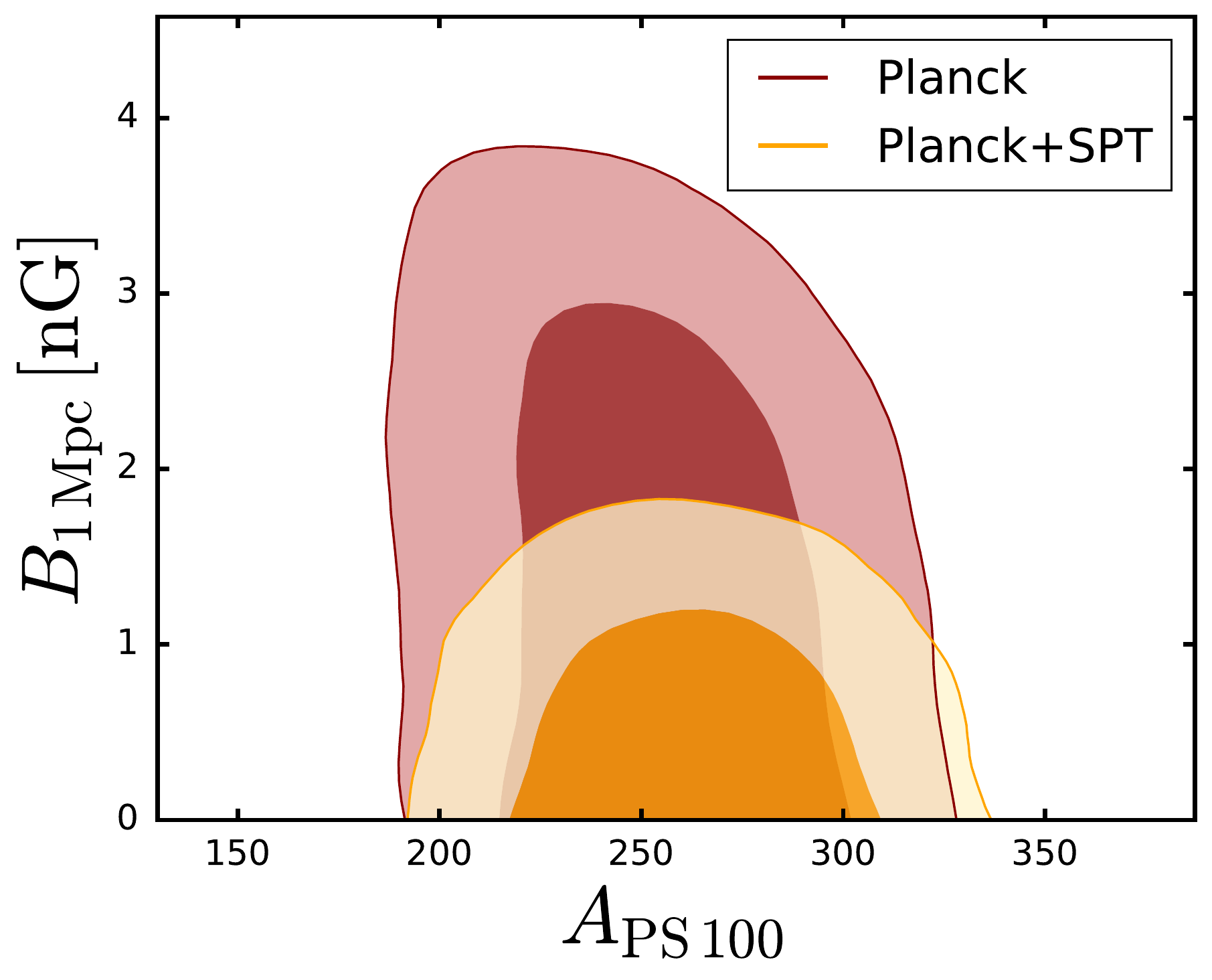}} \,
\subfigure{\includegraphics[width=0.225\textwidth]{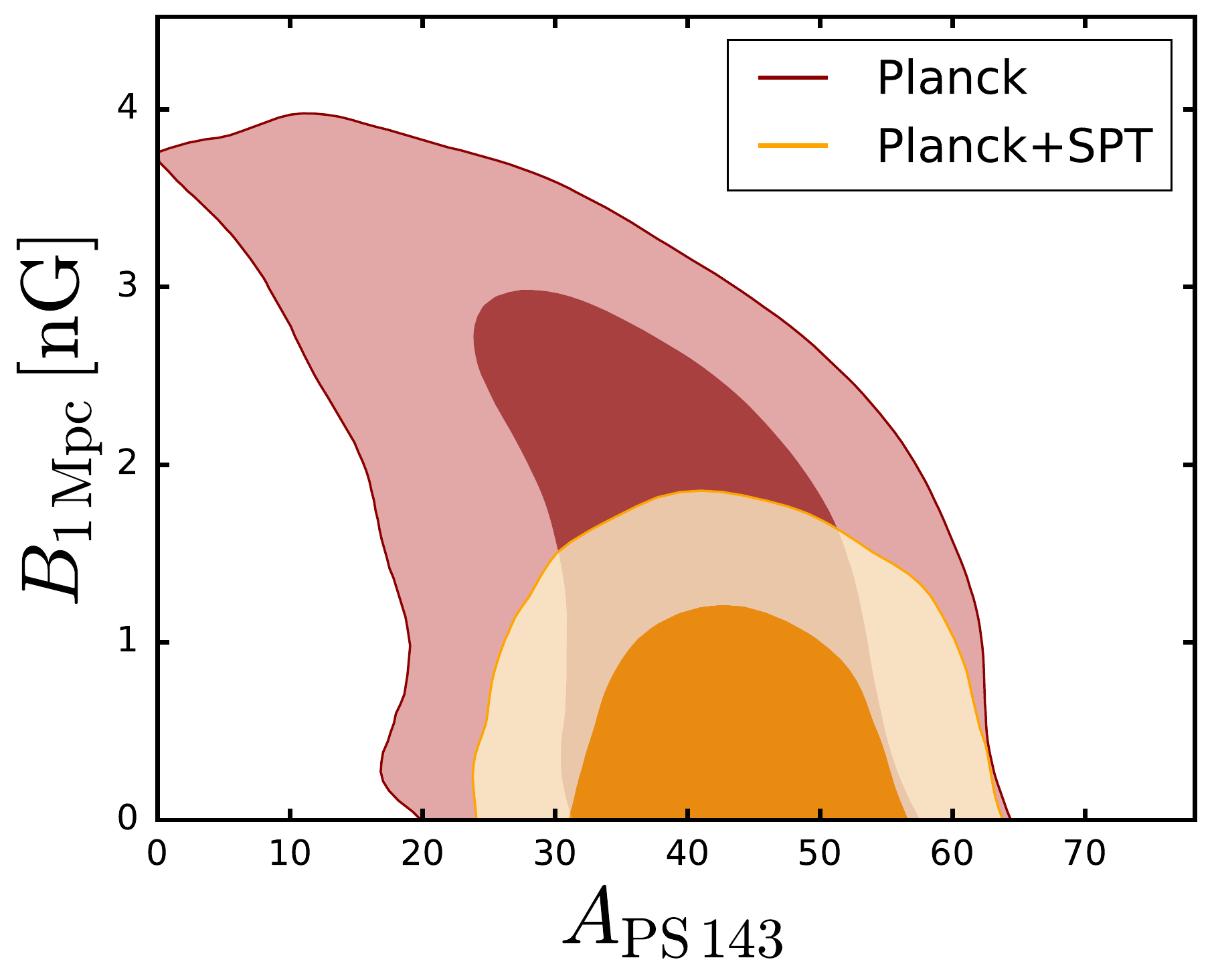}} \\
\subfigure{\includegraphics[width=0.225\textwidth]{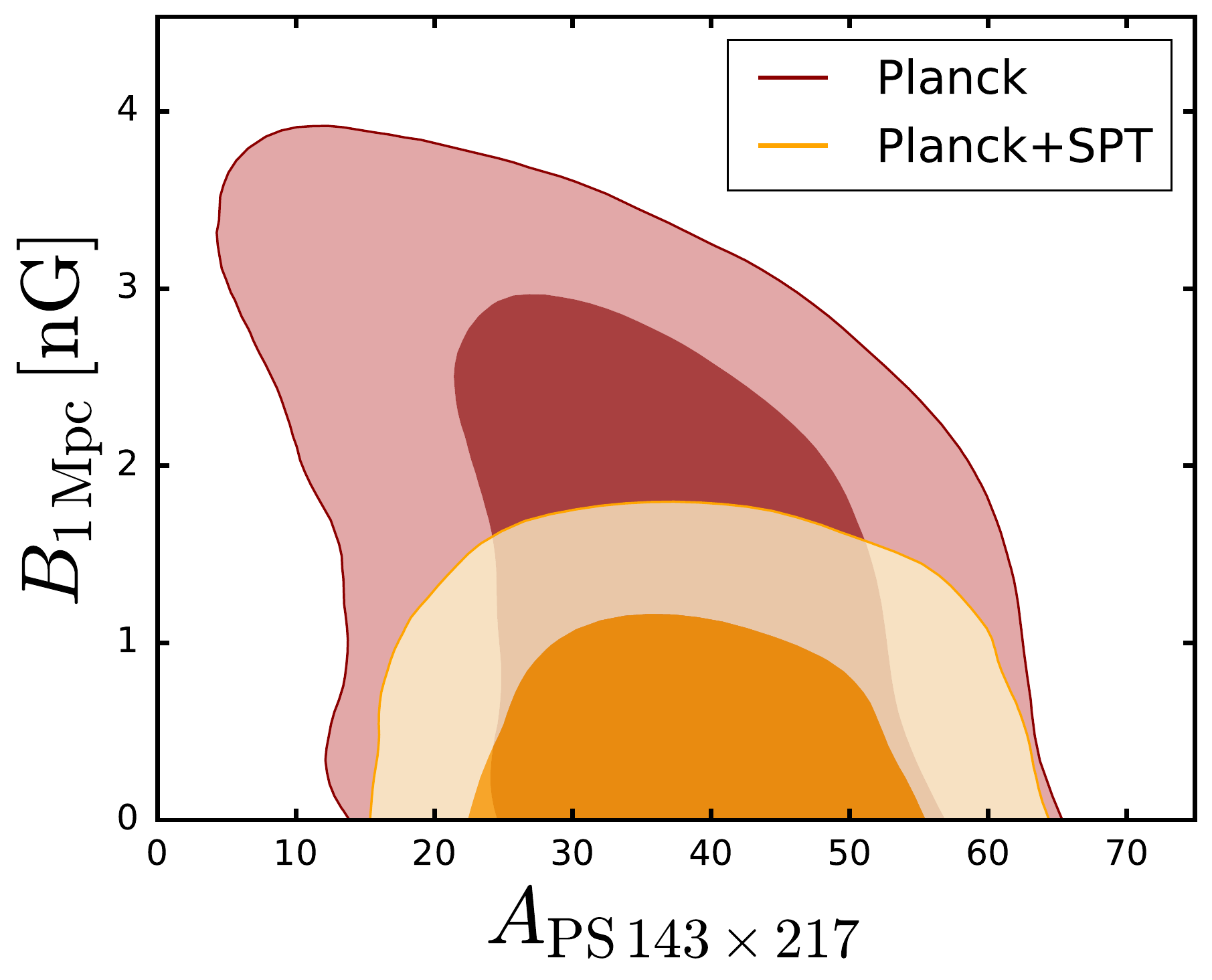}} \,
\subfigure{\includegraphics[width=0.225\textwidth]{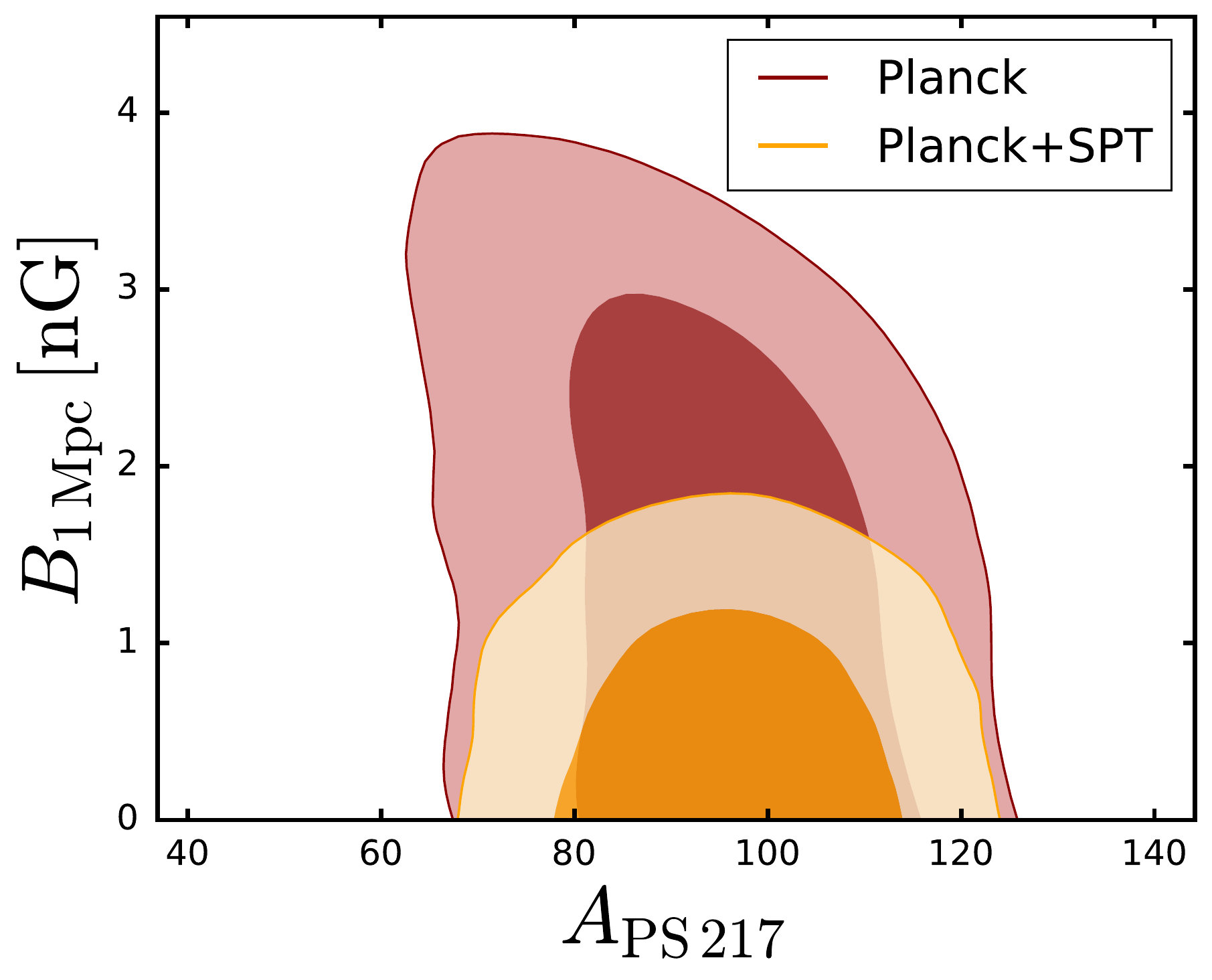}}
\caption{\label{fig:AstrophysicalResidualsPlanck} The 68\% (dark shading) and the 95\% (light shading) CL contours of the joint probabilities for the astrophysical residuals of the \textsc{Planck} likelihood and the PMF amplitude $B_{1 \, \text{Mpc}}$. Adding the \textsc{SPT} B-mode data reduces the degeneracy between the two parameters.}
\end{figure}

\begin{figure}[!htbp]
\centering
\subfigure{\includegraphics[width=0.225\textwidth]{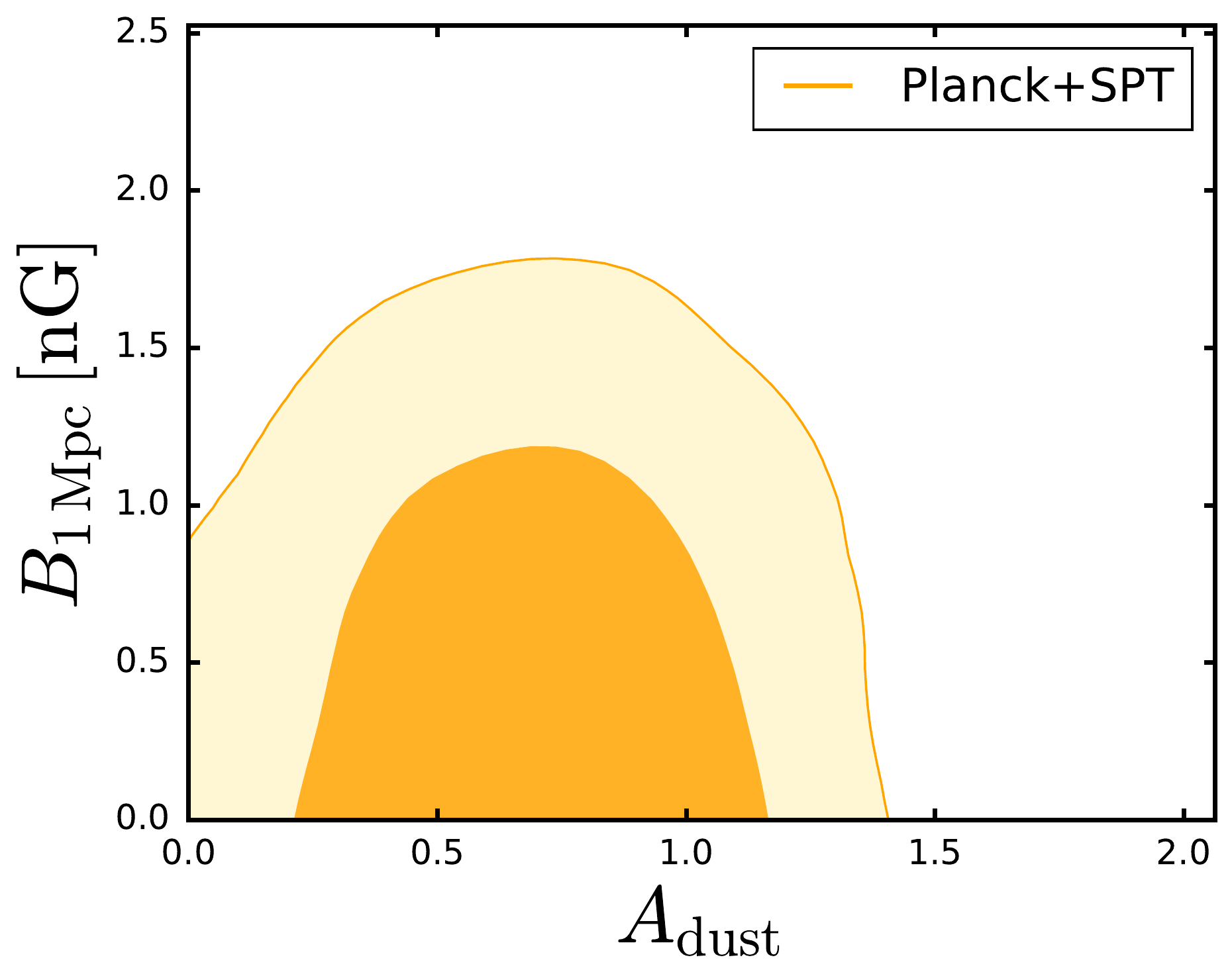}} \,
\subfigure{\includegraphics[width=0.225\textwidth]{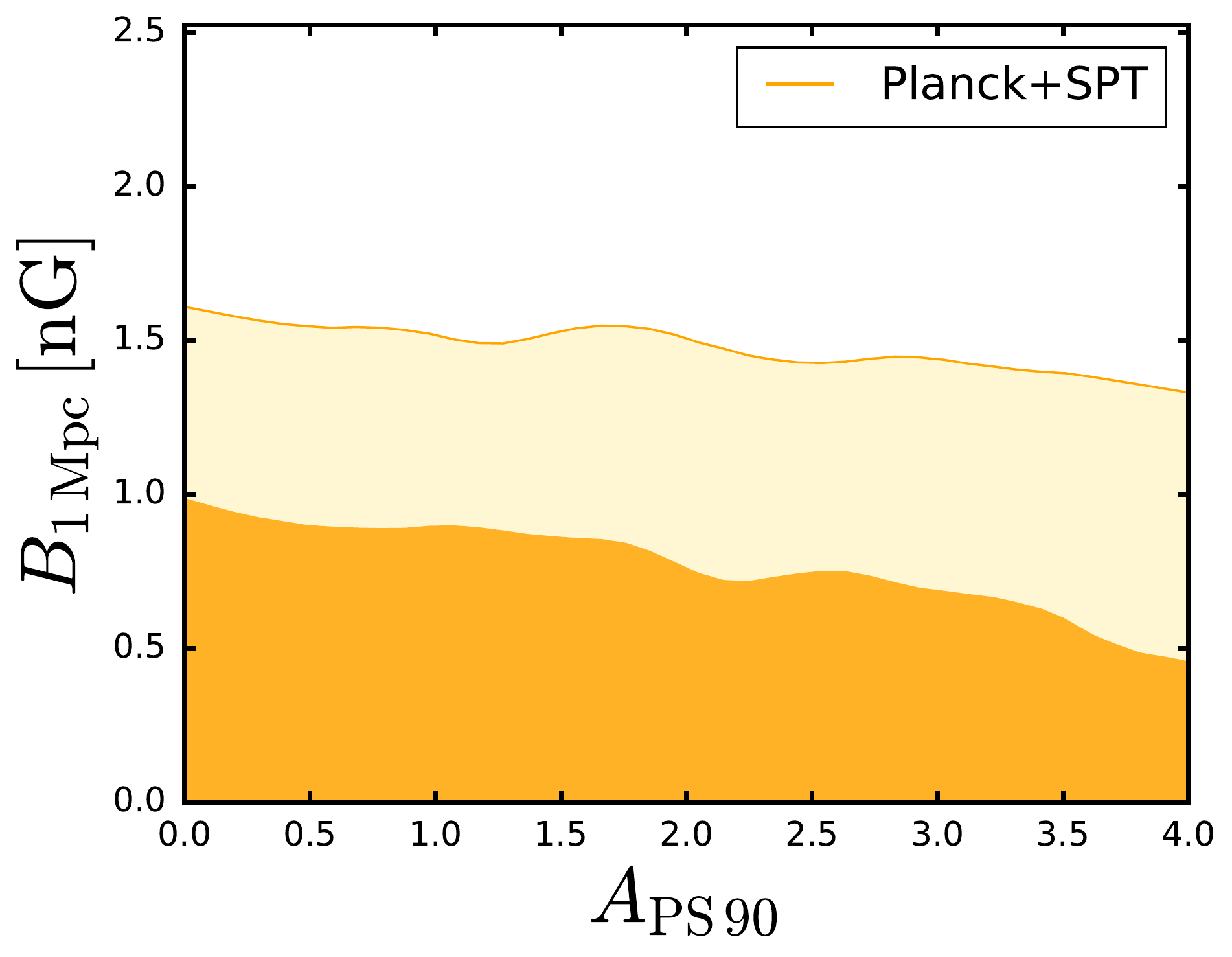}} \\
\subfigure{\includegraphics[width=0.225\textwidth]{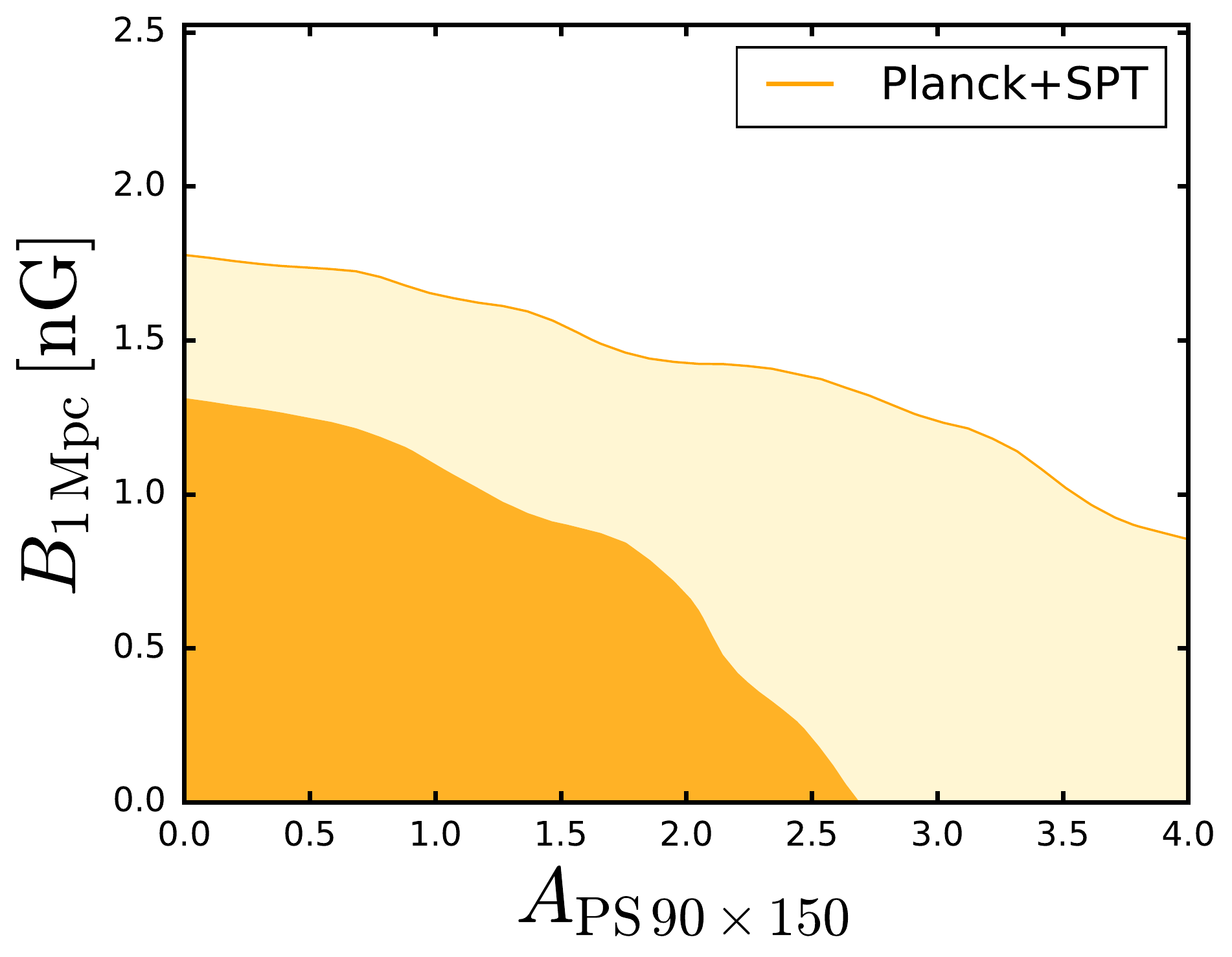}} \,
\subfigure{\includegraphics[width=0.225\textwidth]{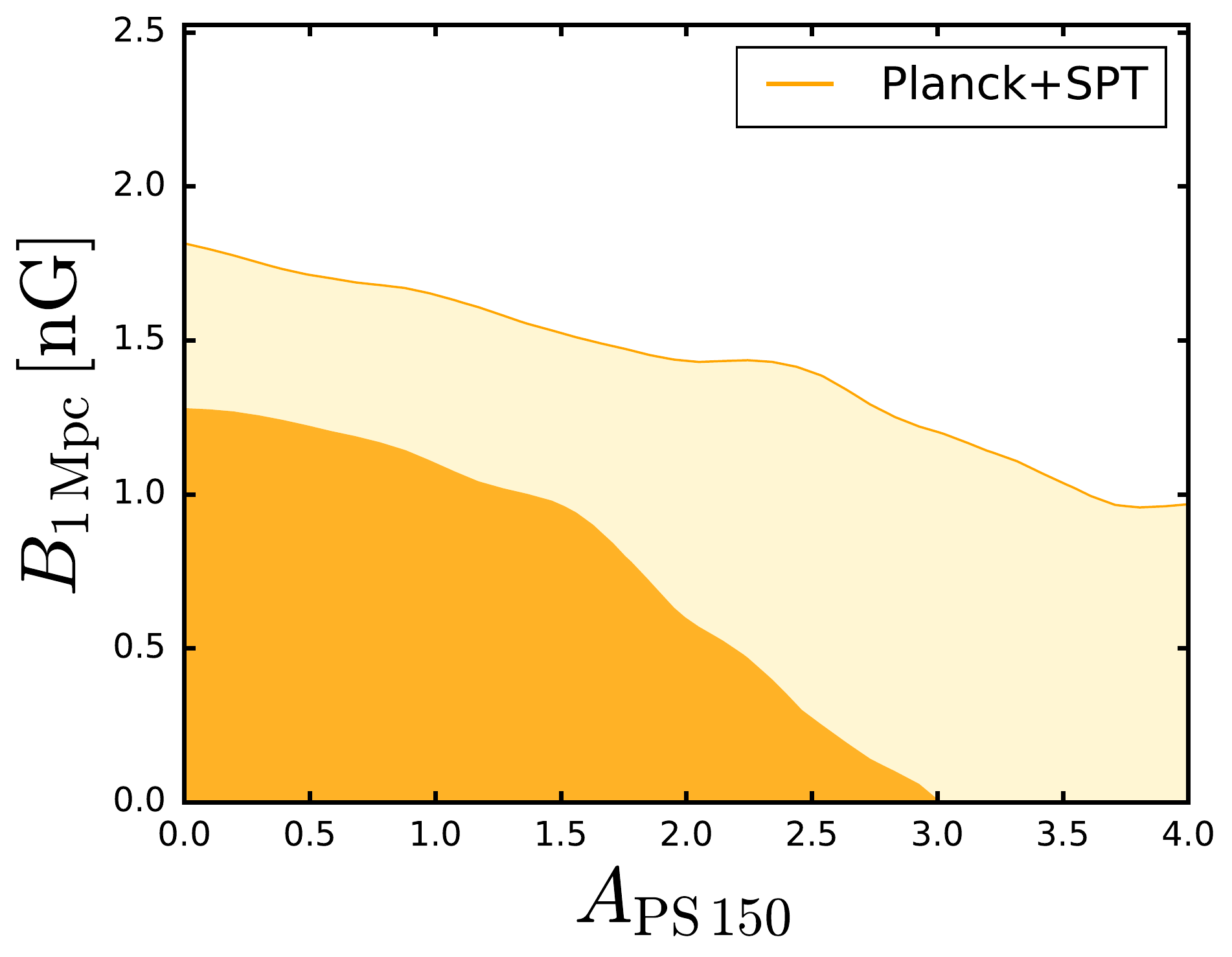}}
\caption{\label{fig:AstrophysicalResidualsSPT} The 68\% (dark shading) and the 95\% (light shading) CL contours of the joint probabilities for the astrophysical residuals of the \textsc{SPT} likelihood and $B_{1 \, \text{Mpc}}$.}
\end{figure}

The \textsc{Planck} and \textsc{SPT} likelihoods account for contributions of astrophysical foregrounds to CMB polarization. They are quantified in terms of parameters $A_{\text{PS} \, 100}$, $A_{\text{PS} \, 143}$, $A_{\text{PS} \, 143 \times 217}$, $A_{\text{PS} \, 217}$ for \textsc{Planck}, and $A_{\text{dust}}$, $A_{\text{PS} \, 90}$, $A_{\text{PS} \, 90 \times 150}$, $A_{\text{PS} \, 150}$ for \textsc{SPT}. In Figs.~\ref{fig:AstrophysicalResidualsPlanck} and \ref{fig:AstrophysicalResidualsSPT} we show the 68\% and the 95\% CL contours of PDFs of the \textsc{Planck} and \textsc{SPT} astrophysical residuals and the magnetic amplitude $B_{1 \, \text{Mpc}}$. These plots show the impact of the foregrounds on diluting the constraints on the PMF, and how a better understanding of the foregrounds can significantly improve the bound on $B_{1 \, \text{Mpc}}$. One can also see from Fig.~\ref{fig:AstrophysicalResidualsSPT} that adding the \textsc{SPT}  data significantly reduces the degeneracy between $B_{1 \, \text{Mpc}}$ and \textsc{Planck}'s astrophysical residuals.

\end{document}